\documentclass{ws-ijmpa}
\usepackage[dvipsnames,svgnames,x11names,table]{xcolor}
\usepackage[T1]{fontenc}
\usepackage[utf8]{inputenc}
\usepackage{amsmath,amssymb,amsfonts} 
\usepackage{graphicx,float,placeins}
\usepackage{array,multirow,longtable,tabu,booktabs,enumitem}
\usepackage{caption,subcaption}
\usepackage{tikz}
\usepackage{tikz-3dplot}
\usetikzlibrary{arrows.meta,shapes.geometric,calc,decorations.pathmorphing,patterns,positioning}
\usepackage[compat=1.0.0]{tikz-feynman}
\usepackage{verbatim,wrapfig,slashed}
\usepackage[super]{cite}
\usepackage[verbose,hypertexnames=false]{hyperref}
\hypersetup{colorlinks=false,allbordercolors=blue,pdfborderstyle={/S/U/W 1}}

\def\r{\boldsymbol{r}}
\def\x{\boldsymbol{x}}
\def\v{\boldsymbol{v}}

\newcommand{\GlauberVertexHard}[1]{%
\begin{tikzpicture}[#1]%
\draw[fill, thick, White] (0,0) circle (5pt); \draw (0,-0.2) edge [line width = 0.8mm, GreenYellow] (0,0.2) node [] {}; \draw[ultra thick, Blue] (0,0) circle (5pt);
\end{tikzpicture}%
}
\newcommand{\GlauberVertexJetA}[1]{%
\begin{tikzpicture}[#1]%
\draw[fill, thick, White] (0,0) circle (5pt); \draw (0,-0.2) edge [line width = 0.8mm, GreenYellow] (0,0.2) node [] {}; \draw[ultra thick, Green] (0,0) circle (5pt);
\end{tikzpicture}%
}
\newcommand{\GlauberVertexJetB}[1]{%
\begin{tikzpicture}[#1]%
\draw[fill, thick, White] (0,0) circle (5pt); \draw (0,-0.2) edge [line width = 0.8mm, GreenYellow] (0,0.2) node [] {}; \draw[ultra thick, teal] (0,0) circle (5pt);
\end{tikzpicture}%
}
\newcommand{\GlauberTextHard}{\raisebox{\dimexpr 2.2ex-\height}{\GlauberVertexHard{}}}
\newcommand{\GlauberTextJetA}{\raisebox{\dimexpr 2.2ex-\height}{\GlauberVertexJetA{}}}
\newcommand{\GlauberTextJetB}{\raisebox{\dimexpr 2.2ex-\height}{\GlauberVertexJetB{}}}

\begin{document}

\markboth{Yao Ma}{Identifying regions for asymptotic expansions of amplitudes}

%
\catchline{}{}{}{}{}
%

\title{Identifying regions for asymptotic expansions of amplitudes: fundamentals and recent advances}

\author{Yao Ma}

\address{Institute for Theoretical Physics, ETH Zürich,\\Zürich 8093, Switzerland\\yaomay@phys.ethz.ch}

\maketitle


\begin{abstract}
This review paper discusses the identification of regions, a crucial first step in applying the ``method-of-regions'' technique. A systematic approach based on Newton polytope geometry has proven successful and efficient for many cases. However, obtaining the correct list of regions becomes increasingly subtle with higher loop numbers or specific Feynman graph topologies. This paper explores the scenarios where such subtleties arise, outlines general strategies to address them, and reviews the current understanding of region structures in various asymptotic expansions of Feynman integrals.

\end{abstract}


\ccode{PACS numbers: 03.65.$-$w, 04.62.+v}

\section{Introduction}
\label{section-introduction}

Evaluating multiloop Feynman integrals remains a central challenge in perturbative quantum field theory, particularly when multiple scales emerge in the external kinematics. A powerful approach to this problem is asymptotic expansion, which systematically disentangles the interplay of disparate scales. Consider a physical process characterized by a small scale $\Lambda_\textup{small}$ (e.g. the mass of a light particle) and a large scale $\Lambda_\textup{large}$ (e.g. the center-of-mass energy in a hard process). By introducing an expansion parameter $\lambda\sim \Lambda_\textup{small}/\Lambda_\textup{large} \ll 1$, the integral $\mathcal{I}(\lambda,\epsilon)$ admits a generalized series expansion of the form:
\begin{eqnarray}
\label{eq:generalized_series_expansion}
    \mathcal{I}(\lambda,\epsilon)\sim \sum_{\mu,k,n} c_{\mu,k,n}(\epsilon)\cdot \lambda^{\mu(\epsilon)}\cdot \textup{log}^k \lambda\cdot \lambda^n
\end{eqnarray}
where $\epsilon$ denotes the dimensional regularization parameter, $\mu(\epsilon)$ linearly depends on $\epsilon$, and the integers $k,n$ track the logarithmic and power-law behaviors of $\lambda$, respectively.~\cite{LeeSmnSmn18} Such expansions not only simplify calculations but also reveal the rich infrared structure of gauge theories by exposing hierarchies between hard, collinear, and soft dynamics that underlie factorization theorems and resummation techniques~\cite{EdenLdshfOlvPkhn02book,Stm95book,Stm96lectures,ClsSprStm04,Cls11book,BurStw13lectures,BchBrgFrl15book,AgwMgnSgnrlTrpth21}.

Among various techniques of doing asymptotic expansions, the one that we focus on in this paper is the ``method of regions'' (MoR)~\cite{BnkSmn97}, which is also referred to as the ``expansion by regions'' or ``strategy of regions''. The basic statement is that, for the original integral $\mathcal{I}(\lambda,\epsilon)$ above, there always exists a certain set of ``regions'' in the entire integration space, $R_1,R_2,\dots,R_n$, such that $\mathcal{I}$ can be restored by summing over expansions around them. Namely,
\begin{eqnarray}
\label{eq:MoR_defining_equation}
    \mathcal{I} = \mathcal{I}^{(R_1)} + \mathcal{I}^{(R_2)} +\dots+ \mathcal{I}^{(R_n)}.
\end{eqnarray}
Note that for each term on the right-hand side, the integrand is modified through a corresponding Taylor expansion, while the integration measure remains as the entire integration space. As an important ingredient for the MoR, we note that evaluating each Feynman integral requires dimensional regularization, and all the scaleless integrals (e.g. $\int d^D k (k^2)^\alpha$) are set to zero.

As an example, the expansion of the one-loop Sudakov form factor, as described in section~2.2 of Ref.~\citen{BchBrgFrl15book}, treats one external momentum $q$ to be the large energy scale, and $p_1$ and $p_2$ to be equally close to distinct lightcones: $p_1^2\sim p_2^2\sim \lambda q^2$ with the expansion parameter $\lambda\ll 1$. In this case, there are four regions in the loop-momentum space $k^\mu = (k^+,k^-,\boldsymbol{k}_\perp)$:
\begin{itemize}
    \item \textup{Hard region: }\quad $k^\mu\sim q(1,1,1)$;
    \item \textup{Collinear-1 region: }\quad $k^\mu\sim q(1,\lambda,\lambda^{1/2})$;
    \item \textup{Collinear-2 region: }\quad $k^\mu\sim q(\lambda,1,\lambda^{1/2})$;
    \item \textup{Soft region: }\quad $k^\mu\sim q(\lambda,\lambda,\lambda)$.
\end{itemize}
By performing the corresponding expansions, the contribution from each region can be expressed in terms of Appell’s hypergeometric functions~\cite{AnstsGlvOlr00}. After summing up these contributions one reproduces the original Feynman integral \emph{to all orders}.

More examples of applying the MoR can be found in papers in the recent 20 years. Although the MoR has appeared as a powerful technique, such that it works for all the examples so far, two fundamental questions regarding its theoretical foundation remain to be answered:
\begin{itemize}
    \item [$\textit{1}$.] Is there a rigorous proof, or a counterexample, of the MoR?
    \item [$\textit{2}$.] Suppose the MoR always works, how to identify the correct list of regions systematically?
\end{itemize}
These two questions are, in fact, closely related. On the one hand, a rigorous answer to question $\textit{1}$ depends on $\textit{2}$: any proof of the MoR must assume a correct choice of regions, while a counterexample of the MoR may simply stem from overlooking or double-counting certain regions. On the other hand, advances in addressing $\textit{1}$ can inform $\textit{2}$. As Jantzen noted in Ref.~\citen{Jtz11}, the MoR is valid to all orders if the expansions satisfy specific conditions, such as domain compatibility, convergence, commutativity, etc. Throughout this paper, we shall focus on answering question $\textit{2}$.

Actually, an all-order identification of regions has been established for Feynman integrals with Euclidean kinematics\cite{Pivovarov:1986xz,Smn90}. A detailed demonstration can be found in Smirnov's book~\citen{Smn02book}, where each region is characterized by a corresponding subgraph, termed the ``asymptotically irreducible'' subgraph. In this sense, expansions around regions can be seen as expansions regarding certain subgraphs.

However, it remains highly nontrivial to extend this statement to Minkowski kinematics, where regions are usually determined based on heuristic examples or experience. This difficulty arises from two key challenges. First, the infrared structure in Minkowski space is more intricate: while in Euclidean spacetime $k^2=0$ requires all components of $k^\mu$ to vanish, in Minkowski spacetime it can also occur when $k^\mu$ lies on a lightcone, corresponding to jet-like configurations in infrared singularities. Second, the mode structure near singularities is complex. For the Sudakov form factor example above, one might wonder why additional modes (e.g., $k^\mu\sim \lambda^2 q$) are not involved in the MoR. A simple answer is that those scalings of $k^\mu$ will lead to scaleless integrals which do not contribute to Eq.~(\ref{eq:MoR_defining_equation}). However, for Feynman graphs at higher loop orders, a more refined and systematic analysis is often required.

This paper reviews systematic approaches to identifying regions in the MoR, with a particular focus on recent advances from Refs.~\citen{GrdHzgJnsMaSchlk22,Ma23,GrdHzgJnsMa24}. In Section~\ref{section-general_remarks_identification_regions}, we make some general remarks on the identification of regions. In particular, regions can be classified into two distinct types: \emph{facet regions} and \emph{hidden regions}. Section~\ref{section-regions_onshell_expansion_wideangle_kinematics} specializes to the on-shell expansion for wide-angle scattering, presenting all-loop results for both region types. These findings are extended in Section~\ref{section-other_asymptotic_expansions_wideangle_scenario} to some other wide-angle expansions, including the \emph{soft expansion}, \emph{timelike-collinear expansion}, and \emph{mass expansion} for heavy-to-light decays. Section~\ref{section-regions_Regge_limit_expansion} examines the \emph{Regge-limit expansion} for the $2\to2$ scattering process, highlighting the role of Glauber singularities in region analysis. We conclude with a summary and outlook in Section~\ref{section-summary_outlook}.

\section{General remarks on identifying regions}
\label{section-general_remarks_identification_regions}

This section establishes the foundation and introduces key concepts for subsequent analyses. In Section~\ref{section-regions_Newton_polytopes}, we review the geometric approach to regions based on Newton polytopes. With this knowledge, Section~\ref{section-facet_regions_hidden_regions} then classifies regions of generic asymptotic expansions into two distinct types, facet regions and hidden regions, whose analyses employ different advanced strategies discussed in Section~\ref{section-basic_strategies}.

\subsection{Regions and Newton polytopes}
\label{section-regions_Newton_polytopes}

For any scalar Feynman integral corresponding to the graph $G$ that is dimensionally regularized in $D=4-2\epsilon$ dimensions, we consider the Lee-Pomeransky representation~\cite{LeePmrsk13}:
\begin{equation}
\mathcal{I}(\boldsymbol{s}) = \mathcal{C}\cdot \int_0^\infty \left( \prod_{e\in G} \frac{dx_e}{x_e}\right)\cdot \left(\prod_{e\in G} x_e^{\nu_e} \right) \cdot \Big( \mathcal{P}(\boldsymbol{x};\boldsymbol{s}) \Big)^{-D/2},
\label{eq:lee_pomeransky_definition}
\end{equation}
where $\mathcal{C}$ is an overall factor, $e$ represents the edges of $G$, $\boldsymbol{x}$ denotes the set of Lee-Pomeransky parameters, and $\boldsymbol{s}$ denotes the set of scalar products formed among the external momenta of $G$. The Lee-Pomeransky polynomial, $\mathcal{P}(\boldsymbol{x};\boldsymbol{s})$, is defined as
\begin{eqnarray}
\label{lee_pomeransky_integrand_definition}
    \mathcal{P}(\boldsymbol{x};\boldsymbol{s})\equiv \mathcal{U}(\boldsymbol{x})+\mathcal{F}(\boldsymbol{x};\boldsymbol{s}),
\end{eqnarray}
where $\mathcal{U}(\boldsymbol{x})$ and $\mathcal{F}(\boldsymbol{x};\boldsymbol{s})$ are the first and second Symanzik polynomials, given by:
\begin{eqnarray}
\mathcal{U}(\boldsymbol{x})=\sum_{T^1}^{}\prod_{e\notin T^1}^{}x_e,\qquad \mathcal{F}(\boldsymbol{x};\boldsymbol{s})=\sum_{T^2}^{} (-s_{T^2}^{}) \prod_{e\notin T^2}^{}x_e +\mathcal{U}(\boldsymbol{x})\sum_{e}^{}m_e^2 x_e\ .
\label{UFterm_general_expression}
\end{eqnarray}
The notations $T^1$ and $T^2$ respectively denote a \emph{spanning tree} and a \emph{spanning 2-tree} of the Feynman graph $G$.

\begin{figure}[h]
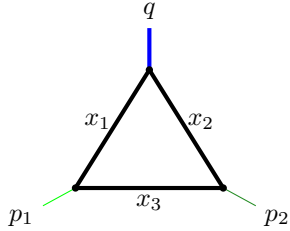

\centering
\include{figs/parameterization_1loop_form_factor}
\vspace{-1em}\caption{Parameterization of the one-loop Sudakov form factor}
\label{figure-parameterization_1loop_form_factor}
\end{figure}

Regions in loop-momentum space, characterized by specific momentum scalings, can be equivalently represented in parameter space through corresponding scalings of (some functions of) the Lee-Pomeransky parameters.\footnote{In general, one can use the relation $x_e\sim D_e^{-1}$, where $D_e$ denotes the denominator corresponding to edge $e$, to translate the momentum scalings to the Lee-Pomeransky parameter scalings.} For the Sudakov form factor example in Section~\ref{section-introduction}, let us parameterize its edges as Fig.~\ref{figure-parameterization_1loop_form_factor}.
Each region (hard, collinear-1, collinear-2, and soft) can then be described as follows:
\begin{itemize}
    \item \textup{Hard region: }\quad $x_1\sim x_2\sim x_3\sim\lambda^0$;
    \item \textup{Collinear-1 region: }\quad $x_1\sim x_3\sim\lambda^{-1},\ x_2\sim\lambda^0$;
    \item \textup{Collinear-2 region: }\quad $x_2\sim x_3\sim\lambda^{-1},\ x_1\sim\lambda^0$;
    \item \textup{Soft region: }\quad $x_1\sim x_2\sim\lambda^{-1},\ x_3\sim\lambda^{-2}$.
\end{itemize}
One key advantage of considering the MoR in parameter space is that it provides a systematic way of identifying the regions~\cite{PakSmn11,JtzSmnSmn12,SmnvSmnSmv19}. In the remaining part of this subsection, we briefly sketch the routine. First, let us consider any $N$-edge Feynman graph $G$ in the Lee-Pomeransky representation~(\ref{eq:lee_pomeransky_definition}), and map each individual term of $\mathcal{P}(\boldsymbol{x};\boldsymbol{s})$ to a specific point in the $(N+1)$-dimensional space as follows:
\begin{eqnarray}
\label{eq:map_LP_term_point}
    s\cdot x_1^{a_1}x_2^{a_2}\cdots x_N^{a_N} \rightarrow (a_1,a_2,\dots,a_N,b)\quad\textup{if }s\sim \lambda^b.
\end{eqnarray}
Next, we construct a Newton polytope $\Delta(\mathcal{P})$, which is defined as the \emph{convex hull} of all the points obtained from (\ref{eq:map_LP_term_point}). Note that by definition, there must be some points on the boundary of $\Delta(\mathcal{P})$ while some may be located in its interior.

\begin{figure}[h]
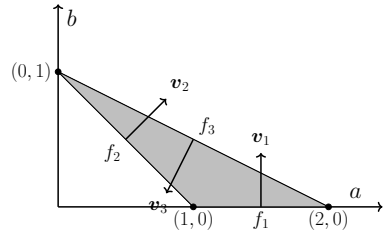

\centering
\include{figs/lower_facet_examples}
\vspace{-2em}\caption{The polytope $\Delta$ (the shaded area) constructed from $\mathcal{P}(x;\lambda)=x^2+x+\lambda$ with lower facets $f_1$ and $f_2$.}
\label{figure-lower_facet_examples}
\end{figure}

Then, a crucial observation is the correspondence between regions and the \emph{lower facets} of $\Delta(\mathcal{P})$. By ``facet'', we refer to those $N$-dimensional faces of $\Delta(\mathcal{P})$, and ``lower facets'' refer to those facets whose inward-pointing normal vectors have positive $(N+1)$-th entries. For example, in Fig.~\ref{figure-lower_facet_examples} we show the polytope constructed from $\mathcal{P}(x;\lambda)= x^2+x+\lambda$. According to~(\ref{eq:map_LP_term_point}), the terms $x^2$, $x$, and $\lambda$, expressed in terms of $x^a\lambda^b$, correspond to the points $(a,b)=(2,0)$, $(1,0)$, and $(0,1)$, respectively. The Newton polytope defined as the convex hull of these points has three facets $f_1$, $f_2$, and $f_3$. By examining the signs of the last (2nd for this example) entries of their inward-pointing normal vectors $\boldsymbol{v}_1$, $\boldsymbol{v}_2$ and $\boldsymbol{v}_3$, we know that $f_1$ and $f_2$ are lower facets, while $f_3$ is not.

For any given lower facet $f_R$ where $R$ is the corresponding region, if we rescale its inward-pointing normal vector such that the last entry is fixed at $1$, the obtained vector $\boldsymbol{v}_R$ is called the \emph{region vector}. In particular, the first $N$ entries of $\boldsymbol{v}_R$ are exactly the exponents of $\lambda$ in the Lee-Pomeransky parameter scaling, in the corresponding region $R$.

Let us again review the Sudakov form factor example for illustration. Here we have $N=3$, thus the Newton polytope is four dimensional, defined as the convex hull of the points $(1,0,0,0)$, $(0,1,0,0)$, $(0,0,1,0)$, $(1,0,1,1)$, $(0,1,1,1)$, and $(1,1,0,0)$, which can be read from the Lee-Pomeransky polynomial
\begin{eqnarray}
    \mathcal{P}(x_1,x_2,x_3;p_1,p_2,q)=x_1+x_2+x_3+(-p_1^2)x_1x_3+(-p_2^2)x_2x_3+(-q^2)x_1x_2.\nonumber
\end{eqnarray}
One can further check that this polytope has 4 distinct lower facets, with the corresponding region vectors:
\begin{itemize}
    \item $\boldsymbol{v}_H=(0,0,0,1)$\quad $\rightarrow$\quad \textup{hard region vector};
    \item $\boldsymbol{v}_{C_1}=(-1,0,-1,1)$\quad $\rightarrow$\quad \textup{collinear-1 region vector};
    \item $\boldsymbol{v}_{C_2}=(0,-1,-1,1)$\quad $\rightarrow$\quad \textup{collinear-2 region vector};
    \item $\boldsymbol{v}_{S}=(-1,-1,-2,1)$\quad $\rightarrow$\quad \textup{soft region vector}.
\end{itemize}

In essence, this geometric approach to the MoR translates the task of identifying relevant regions of a Feynman integral, into the task of identifying the lower facets of the corresponding Newton polytope. Based on this approach, computer codes such as \texttt{Asy2}~\cite{PakSmn11,JtzSmnSmn12}, \texttt{ASPIRE}~\cite{AnthnrySkrRmn19}, \texttt{pySecDec}~\cite{HrchJnsSlk22}, \texttt{AmpRed}~\cite{Chen25AmpRed}, etc., have been developed.

We conclude this subsection by highlighting two major challenges addressed in the following sections.

First, while existing computer codes have proven effective provided the number of edges is not too large, it is natural to raise the following question: \emph{for a given expansion of interest, can one establish a general rule, which governs all the regions and specifies all the relevant modes for arbitrary Feynman graphs?} The answer to this question is particularly valuable when one aims at an all-loop-order statement of regions, or when a finite but large number of edges are involved, thus being associated with a high-dimensional Newton polytope, for which the aforementioned computer codes may struggle to produce timely results. This question, therefore, delves into the Newton polytope approach, and its resolution surpasses the capabilities of current computer codes.

Second, the geometric approach may fail to identify those regions ``hidden'' inside $\Delta(\mathcal{P})$, which are referred to as \emph{hidden regions}~\cite{GrdHzgJnsMa24}. Actually, this has already been noticed in Ref.~\citen{JtzSmnSmn12}, where, in order to solve this issue, the code \texttt{Asy} has been upgraded to \texttt{Asy2}. However, as we will demonstrate later, \texttt{Asy2} is only able to identify those hidden regions with relatively simple $\mathcal{F}$ polynomial structures. In general, hidden regions typically emerge when the $\mathcal{F}$ polynomial contains terms with opposite signs whose cancellations satisfy the Landau equations~\cite{Lnd59}. While this leads to a constrained yet still complex structure of the $\mathcal{F}$ polynomial, it often corresponds to remarkably elegant features in the underlying Feynman graph topology, as we shall specify later.

\subsection{Facet regions and hidden regions}
\label{section-facet_regions_hidden_regions}

The main focus of this review paper is to address the two issues above. To this end, we classify regions into two types: facet regions and hidden regions. Roughly speaking, for a given expansion, its associated facet regions are those corresponding to the lower facets of $\Delta(\mathcal{P})$, while its hidden regions are those contained within $\Delta(\mathcal{P})$. In other words, this classification is based on whether regions are identifiable by the geometric approach.

This defining property manifests in their distinct parameter-space representations. As previously discussed, regions can be characterized in parameter space via scalings of functions derived from the Lee-Pomeransky parameters. For facet regions, these functions reduce to the parameters themselves. Namely, relations of the form $x_i\sim \lambda^{v_i}$ ($i=1,\dots,N$) are sufficient to represent any given facet region, where the value of $x_i$ can be chosen freely at $\mathcal{O}(\lambda^{v_i})$. By substituting the scalings of $x_1,\dots,x_N$ into $\mathcal{F}(\boldsymbol{x};\boldsymbol{s})$, one can obtain the scaling of each individual term. The dominant terms constitute the \emph{leading $\mathcal{F}$ polynomial}, which is responsible for the leading-order expansion around this region.

All four regions in the form factor example (hard, collinear-1, collinear-2, soft) are facet regions. More intricate cases arise in some very recent works~\citen{JskwJnsSzfUrch25,HouWangZhang25}, where all regions are likewise facet regions. (Note that the region vectors in those papers omit the last entry $1$.) Later in Section~\ref{section-facet_regions_all_order_result}, we will characterize the facet regions in the on-shell expansion to all loop orders.

Hidden regions, in contrast, cannot be simply described by the scaling relation above. While the magnitudes of the parameters can be described by $x_i\sim \mathcal{O}(\lambda^{v_i})$ with $i=1,\dots,N$, not all of $x_1,\dots,x_N$ can be chosen freely within these ranges; additional constraints imposed by scaling relations further restrict their values. When these parameter magnitudes are substituted into $\mathcal{F}(\boldsymbol{x};\boldsymbol{s})$, the dominant contributions are termed the \emph{super-leading $\mathcal{F}$ terms} for this region. Crucially, individual super-leading terms are larger than their collective sum, which is subject to additional suppression due to the scaling relations. This suppressed sum becomes comparable in magnitude to the leading $\mathcal{F}$ terms. Therefore, both super-leading and leading terms must be retained in the leading-order expansion.

\begin{figure}[h]
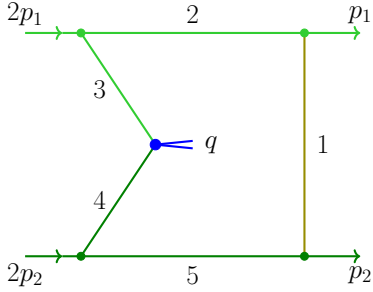
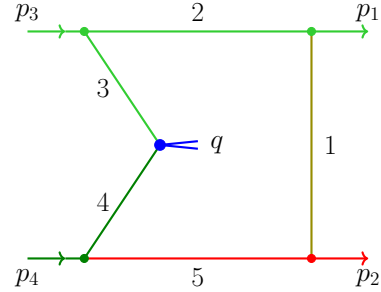

\centering
\hspace{-1em}
\begin{subfigure}[b]{0.4\textwidth}
\include{figs/hidden_region_example1}
\vspace{-2em}
\captionsetup{width=1.05\linewidth}\caption{The unique hidden region of the expansion in Eq.~(\ref{eq:hidden_region_example1_kinematics}), where a Glauber-mode momentum $(\lambda,\lambda,\lambda^{1/2})$ is exchanged between $p_1$ and $p_2$.}
\label{figure-hidden_region_example1}
\end{subfigure}
\hspace{6em}
\begin{subfigure}[b]{0.4\textwidth}
\include{figs/hidden_region_example2}
\vspace{-2em}
\caption{The unique hidden region of the expansion in Eq.~(\ref{eq:hidden_region_example2_kinematics}), where a Glauber-mode momentum $(\lambda,\lambda^2,\lambda)$ is exchanged between $p_1$ and $p_2$.}
\label{figure-hidden_region_example2}
\end{subfigure}
\caption{Examples of hidden regions at one-loop level.}
\label{figure-hidden_region_examples}
\end{figure}

Let us now explain the mechanism above through two pedagogical examples. The first is the one-loop five-point integral discussed in Ref.~\citen{JtzSmnSmn12}, with a particular choice of external kinematics (see Fig.~\ref{figure-hidden_region_example1}):
\begin{eqnarray}
\label{eq:hidden_region_example1_kinematics}
    p_1^2=p_2^2=m_2^2=m_3^2=m_4^2=m_5^2=0,\quad m_1^2\sim \lambda q^2\quad (\lambda\ll1).
\end{eqnarray}
Under this kinematic configuration, the $\mathcal{F}$ polynomial reads
\begin{eqnarray}
    \mathcal{F}(\boldsymbol{x};\boldsymbol{s})= m_1^2x_1(x_1+\dots+x_5) + q^2(x_2-x_3)(x_4-x_5).
\end{eqnarray}
For the asymptotic expansion of $\lambda$, one can obtain the facet regions directly from the geometric approach. In addition, there is a hidden region characterized by either of the following parameter scalings
\begin{subequations}
\label{eq:hidden_region_example1_scaling}
    \begin{align}
        x_1\sim x_2\sim x_3\sim x_4\sim\lambda^{-1},\quad x_5-x_4\sim \lambda^{0};
        \label{eq:hidden_region_example1_scaling_possibility1}\\
        x_1\sim x_2\sim x_4\sim x_5\sim\lambda^{-1},\quad x_3-x_2\sim \lambda^{0}.
        \label{eq:hidden_region_example1_scaling_possibility2}
    \end{align}
\end{subequations}
In momentum space, both scalings above correspond to the Glauber region (Fig.~\ref{figure-hidden_region_example1}), where the momentum carried by edge $1$ scales as $k^\mu=(k^+,k^-,\boldsymbol{k}_\perp)\sim (\lambda,\lambda,\lambda^{1/2})$. In this lightcone coordinate system, we have chosen $p_1^\mu$ to be along the $+$ direction while $p_2^\mu$ the $-$ direction. We note that in either Eq.~(\ref{eq:hidden_region_example1_scaling_possibility1}) or (\ref{eq:hidden_region_example1_scaling_possibility2}), the super-leading $\mathcal{F}$ terms are given by $q^2(x_2-x_3)(x_4-x_5)$, which individually scale as $\lambda^{-2}$, while suppressed to $\lambda^{-1}$ in the sum.

Another example of hidden region, which is slightly more complicated, is discussed in a recent work~\citen{BchHgrJskwNbtSwbch24}. The Feynman graph topology is the same as above (one-loop five-point graph), with a different external kinematics (see Fig.~\ref{figure-hidden_region_example2}):
\begin{eqnarray}
\label{eq:hidden_region_example2_kinematics}
    p_1^2=p_2^2=p_3^2=p_4^2=m^2=0,\quad p_1\cdot p_3\sim \lambda^2 q^2,\quad p_2\sim\lambda q.
\end{eqnarray}
One crucial difference from Eq.~(\ref{eq:hidden_region_example1_kinematics}) is that the external momentum $p_2$ is soft rather than lightlike. For this kinematics, there is again a hidden region where the momentum carried by edge $1$ scales as $k^\mu=(k^+,k^-,\boldsymbol{k}_\perp)\sim (\lambda,\lambda^2,\lambda)$, where we have taken $p_1^\mu$ and $p_3^\mu$ to be along the $+$ direction, and $p_2^\mu$ the $-$ direction. In comparison with the previous example, the asymmetry in this Glauber scaling is due to the asymmetry of the external momenta. One can also view this hidden region in parameter space, where $x_1,x_2,x_3,x_5$ are $\mathcal{O}(\lambda^{-2})$ while $x_4$ is $\mathcal{O}(\lambda^{-1})$. It is straightforward to check that each super-leading $\mathcal{F}$ term is $\mathcal{O}(\lambda^{-3})$, and their sum is suppressed into $\mathcal{O}(\lambda^{-2})$. More details are provided in Ref.~\citen{BchHgrJskwNbtSwbch24}.

From the examples above, it is natural to deduce a graphical feature which distinguishes facet regions and hidden regions. Let us consider all the connected subgraphs of $G$, for each of which an off-shell momentum (either hard or Glauber) is exchanged. By denoting the union of these subgraphs as $\gamma_\textup{off}^{}$, one can check that $\gamma_\textup{off}^{}$ is connected for facet regions, while disconnected for hidden regions. To be explicit, for both examples in Fig.~\ref{figure-hidden_region_examples}, $\gamma_\textup{off}^{}$ consists of a hard vertex (attached by $q^\mu$) and a Glauber-mode edge (the one connecting $p_1^\mu$ and $p_2^\mu$), which is clearly disconnected. We will return to this striking feature later.

To summarize, Table~\ref{table-facet_region_hidden_region} lists some properties of facet regions and hidden regions.
\begin{table}[h]
\tbl{Some properties of facet and hidden regions.\label{table-facet_region_hidden_region}}
{\begin{tabular}{@{}l l l@{}} \toprule
 & {\bf Facet regions} & {\bf Hidden regions} \\ \colrule
Geometric location & lower facets of $\Delta(\mathcal{P})$. & interior of $\Delta(\mathcal{P})$. \\ \addlinespace
\multirow{2}{*}{Parameter scaling} & \multirow{2}{*}{$x_i \sim \lambda^{v_i}$.} & $x_i = \mathcal{O}(\lambda^{v_i})$ with additional scaling \\
 & & constraints. \\ \addlinespace
\multirow{2}{*}{Leading behavior} 
 & \multirow{2}{*}{leading $\mathcal{U}$ and $\mathcal{F}$ terms.} & super-leading $\mathcal{F}$ terms (sum suppressed due to \\
 & & cancellations) $+$ leading $\mathcal{U}$ and $\mathcal{F}$ terms. \\ \addlinespace
Subgraph structure & $\gamma_\text{off}^{}$ is connected. & $\gamma_\text{off}^{}$ is disconnected. \\ \botrule
\end{tabular}}
\end{table}

\subsection{Basic strategies}
\label{section-basic_strategies}

Our ultimate objective is to develop an all-loop-order description of both facet and hidden regions for arbitrary asymptotic expansions. Below we outline key strategies toward this goal.

\subsubsection{Identifying facet regions via graph-theoretical approaches}
\label{section-identifying_facet_regions}

From Eq.~(\ref{UFterm_general_expression}), each term of $\mathcal{P}(\x;\boldsymbol{s})$ is characterized by either a \emph{spanning tree} or \emph{spanning 2-tree} of the graph~$G$. Moreover, for any region $R$, we can associate each spanning (2-)tree with a certain \emph{weight}, such that the leading terms correspond to \emph{minimum spanning (2-)trees}. To this end, we first associate each edge $e\in G$ with a real number $w(e)$, defined as the exponent of $\lambda$ in the scaling of $x_e$:
\begin{eqnarray}
\label{eq:edge_weights_definition}
    w(e)=v_{R,e},\quad\textup{where }x_e\sim\lambda^{v_{R,e}}\textup{ in region }R.
\end{eqnarray}
For a spanning tree $T^1$ or a spanning 2-tree $T^2$, we then define $w(T^1)$ as the sum over $w(e)$ where $e$ is \emph{not} in $T^1$, and $w(T^2)$ as the sum over $w(e)$ where $e$ is \emph{not} in $T^2$, plus a kinematic contribution $n$ if the kinematic factor $s_{T^2}\sim \lambda^n$. Namely,
\begin{subequations}
    \begin{align}
        & w(T^1)\equiv \sum_{e\notin T^1} w(e);
        \label{eq:definition_spanning_tree_weight}\\
        & w(T^2)\equiv \sum_{e\notin T^2} w(e) + n\quad \text{ if }\ \  s_{T^2}\sim \lambda^n Q^2.
        \label{eq:definition_spanning_2_tree_weight}
    \end{align}
\end{subequations}
Among all the $T^1$ and $T^2$, the minimum spanning (2-)trees are those with the minimum weight. By definition, they correspond to the leading $\mathcal{U}$ and $\mathcal{F}$ terms in the expansion w.r.t. $R$. Consequently, the problem of identifying the facet regions, which is equivalent to identifying the lower facets of $\Delta(\mathcal{P})$, can be reformulated as follows: \emph{we aim to find all the possible scalings of the Lee-Pomeransky parameters $x_1,\dots,x_N$, such that the points, which are associated with the minimum spanning (2-)trees of $G$, span a lower facet of the Newton polytope $\Delta(\mathcal{P})$.}

All-loop-order analyses of facet regions using this graph-theoretical approach present significant challenges but remain achievable. Ref.~\citen{Ma23} demonstrates this for the \emph{on-shell expansion}, where a necessary and sufficient condition for facet regions is established to all loops and summarized as the \emph{on-shell-expansion region theorem}.

We will discuss this result in more detail in Section~\ref{section-facet_regions_all_order_result}. For now, let us briefly outline the basic idea, which is based on the following two criteria.
\begin{itemize}[leftmargin=*,labelsep=0.5em]
    \item {\bf Minimum-weight criterion}: given any leading term $\x^{\r}$ of $\mathcal{P}(\x;\boldsymbol{s})$, the weight $w(T(\r))\leqslant w(T(\r'))$ for any other term $\x^{\r'}$.\footnote{The notation $\x^{\r}$ represents the term $x_1^{r_1}x_2^{r_2}\cdots x_N^{r_N}$ up to its overall coefficient, and $T(\r)$ represents its corresponding spanning (2-)tree.}
    \item {\bf Facet criterion}: for each region $R$, the region vector $\v_R$ is the unique vector normal to $f_R$ (up to rescaling).
\end{itemize}
These criteria are straightforward to justify from their definitions, but they play a central role in the proof. First, many of the ``pathological'' scalings of $x_1,\dots,x_N$ can be excluded by the minimum-weight criterion. That is, for any such scaling, one can always find a leading term $\x^{\r}$ and another term $\x^{\r'}$ such that $w(T(\r)) > w(T(\r'))$. All remaining parameter scalings correspond to momentum-conserving configurations. Additional constraints on the scalings are needed due to the facet criterion, otherwise there would be an extra homogeneity property of the leading polynomial, implying that its dominant terms lie on a face of dimension less than $N$, leading to a scaleless integral. By systematically employing such proof-by-contradiction arguments, one ultimately finds that all valid parameter scalings take the form
\begin{eqnarray}
\label{eq:parameter_scaling_onshell_expansion_region_theorem}
    x_H^{}\sim\lambda^0,\quad x_J^{}\sim\lambda^{-1},\quad x_S^{}\sim\lambda^{-2},
\end{eqnarray}
where $x_H^{}$, $x_J^{}$, and $x_S^{}$ denote parameters associated with the hard ($H$), jet ($J$), and soft ($S$) subgraphs, respectively. The structure of these subgraphs aligns with the infrared picture (Fig.~\ref{figure-onshell_generic_facet_region}, as we will explain later).

We note that the list of facet regions is acutely sensitive to the specific form of the asymptotic expansion. Consider for example an expansion where external momentum scalar products divide into $\{ s_1,s_2,\dots \}\sim \mathcal{O}(1)$ and $\{ t_1,t_2,\dots \}\sim \mathcal{O}(\lambda)$. The complete set of facet regions derived from the above methodology will change under kinematic modifications, e.g., if we set $t_1=0$ while keeping the rest $\{t_2,\dots\}$ still at $\mathcal{O}(\lambda)$. We will see this by comparing the on-shell expansion results and those from other similar expansions in Section~\ref{section-other_asymptotic_expansions_wideangle_scenario}, where additional subgraph constraints needed for soft expansion are explained.

\subsubsection{Identifying hidden regions via dissections of polytopes}
\label{section-identifying_hidden_regions}

In Section~\ref{section-facet_regions_hidden_regions}, we established the distinction between facet and hidden regions through their parametric representations. This fundamentally originates from the difference in their corresponding infrared singularities in parameter space. Generally, each region (except the hard region) corresponds to specific integrand-level singularities: \emph{facet regions are associated with endpoint singularities of the integration measure, while hidden regions arise from singularities located inside the integration domain}. To identify hidden regions in a Feynman integral, one must therefore first determine whether such interior singularities are present.

The Landau equations provide necessary conditions for such integrand-level singularities. In Feynman parameter space, these take the form~\cite{EdenLdshfOlvPkhn02book}:
\begin{eqnarray}
\label{eq:Landau_equations_parameter_space}
    \mathcal{F}=0,\quad\alpha_e \frac{\partial \mathcal{F}}{\partial \alpha_e} = 0\quad \forall e\in G,
\end{eqnarray}
where $\alpha_1,\dots,\alpha_N$ are the Feynman parameters. Note that the first equation $\mathcal{F}=0$ follows directly from the second due to the homogeneity of $\mathcal{F}$.

The existence of pinch singularities in parameter space within the domain of integration must be characterized by a solution of the Landau equations~(\ref{eq:Landau_equations_parameter_space}) where some $\alpha_e$ is positive. This leads to the following requirement: \emph{there exists at least one Feynman parameter $\alpha_e \neq 0$ with $\frac{\partial\mathcal{F}}{\partial \alpha_e}$ containing non-vanishing terms of both signs that cancel on the solution.}

We note that this requirement is satisfied in the motivating examples in Section~\ref{section-facet_regions_hidden_regions}. For example, the Lee-Pomeransky parameter scalings in the hidden region described by Fig.~\ref{figure-hidden_region_example1} are shown in~(\ref{eq:hidden_region_example1_scaling}). In terms of Feynman parameters, we have\footnote{To derive (\ref{eq:hidden_region_example1_scaling_Feynman}) from (\ref{eq:hidden_region_example1_scaling}), one can use the relation $\alpha_i=\frac{x_i}{x_1+\dots+x_N}$ and $\alpha_1+\dots+\alpha_N=1$.}
\begin{subequations}
\label{eq:hidden_region_example1_scaling_Feynman}
    \begin{align}
        \alpha_1\sim \alpha_2\sim \alpha_3\sim \alpha_4\sim\lambda^{0},\quad \alpha_5-\alpha_4\sim \lambda^{1};
        \label{eq:hidden_region_example1_scaling_possibility1_Feynman}\\
        \alpha_1\sim \alpha_2\sim \alpha_4\sim \alpha_5\sim\lambda^{0},\quad \alpha_3-\alpha_2\sim \lambda^{1}.
        \label{eq:hidden_region_example1_scaling_possibility2_Feynman}
    \end{align}
\end{subequations}
Given $\mathcal{F}(\boldsymbol{\alpha};\boldsymbol{s})= m_1^2\alpha_1(\alpha_1+\dots+\alpha_5) + q^2(\alpha_2-\alpha_3)(\alpha_4-\alpha_5)$, it is clear that for the relation (\ref{eq:hidden_region_example1_scaling_possibility1_Feynman}), we have $\alpha_2,\alpha_3\neq 0$ while $\frac{\partial\mathcal{F}}{\partial \alpha_2}$ and $\frac{\partial\mathcal{F}}{\partial \alpha_3}$ contain non-vanishing terms of both signs canceling as $\lambda\to0$. For (\ref{eq:hidden_region_example1_scaling_possibility2_Feynman}) we have $\alpha_4,\alpha_5$ instead.

Building on these insights, we now construct a search algorithm to identify graphs that potentially contain pinch singularities in parameter space. The procedure begins by extracting all non-vanishing $\mathcal{F}$ terms in the kinematic limit and grouping them by independent Mandelstam variables, yielding a set of polynomials $\{\mathcal{F}^{(s_{ij})}\}$, each of which depends on a single Mandelstam variable $s_{ij}$ from the independent set $\{s_{ij}\}$.

\begin{figure}[h]
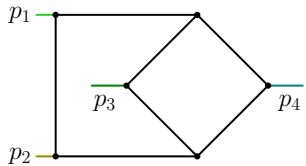

\centering
\include{figs/nonplanr_double_box}
\vspace{-2em}\caption{The nonplanar double-box graph, where all the four momenta are nearly on-shell, denoted by $p_i$ ($i=1,2,3,4$). Although there are positive and negative kinematic invariants in $\mathcal{F}$ for this specific example, no cancellation can be consistent with the Landau equations, thus there are no hidden regions.}
\label{figure-nonplanr_double_box}
\end{figure}

For each $\mathcal{F}^{(s_{ij})}$ (denoted simply as $\mathcal{F}$ in the steps below), we implement the following recursive procedure:
\begin{itemize}
    \item [Step 1.] Compute its positive- and negative-definite parts $\mathcal{F}_+$ and $\mathcal{F}_-$. If either of them vanishes, exit the algorithm outputting that there are no pinch Landau singularities. Otherwise go to Step 2.
    \item [Step 2.] Compute $\partial\mathcal{F}_+/\partial \alpha_e$ and $\partial\mathcal{F}_-/\partial \alpha_e$ for all the $\alpha_e$ that $\mathcal{F}$ depends on. If none of these derivatives vanish, exit the algorithm outputting that there are potential pinch singularities. Otherwise go to Step 3.
    \item [Step 3.] Identify the $\alpha_e$ for which~$\partial\mathcal{F}_+/\partial \alpha_e = 0$ or $\partial\mathcal{F}_-/\partial \alpha_e = 0$, replace $\mathcal{F}$ by $\mathcal{F}|_{\alpha_e=0}$, and return to Step 1.
\end{itemize}
Note that the ``potential pinch singularities'' output represents only a \emph{necessary condition}, as cancellations might occur for unphysical (negative or complex) $\alpha_e$ values, which we exclude from consideration here. Nevertheless, this algorithm suffices to tell quite a few nontrivial graphs without pinch singularities. One example is the nonplanar double-box graph for massless scattering, as shown in Fig.~\ref{figure-nonplanr_double_box}, where the $\mathcal{F}$ polynomial contains terms with coefficients $s$, $t$, and $u$ simultaneously. By using the constraint $u=-s-t$, the coefficients in this $\mathcal{F}$ are necessarily of different signs. However, as analyzed in Appendix A of Ref.~\citen{GrdHzgJnsMaSchlk22}, no cancellations among the $\mathcal{F}$ terms can be consistent with the Landau equations, so there cannot be any hidden regions. The algorithm above carries out the same analysis and outputs ``there are no pinch Landau singularities''.

For those graphs featuring pinch singularities in parameter space, it remains to answer what to do with the hidden regions. Our strategy is to dissect the original polytope $\Delta(\mathcal{P})$ into several sub-polytopes, each representing a certain change of variables from the Lee-Pomeransky parameters. These changes of variables map the original pinch singularities to integration endpoints in the new parameter space, thereby converting hidden regions in $\Delta$ into facet regions of the obtained sub-polytopes. One can then use existing computer codes to analyze the facet regions of these sub-polytopes and obtain a complete set of regions.

Further details on the polytope dissection procedure can be found in Sections 2 and 3 of Ref.~\citen{GrdHzgJnsMa24}. In what follows in this paper, we demonstrate the methodology of identifying regions through some representative examples of asymptotic expansions.

\section{Regions in the on-shell expansion for wide-angle kinematics}
\label{section-regions_onshell_expansion_wideangle_kinematics}

\begin{figure}[h]
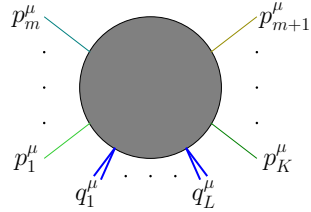

\centering
\include{figs/generic_onshell_expansion}
\vspace{-2em}\caption{Wide-angle scattering with external momenta $\{p_i^\mu\}_{i=1,\dots,K}$ and $\{q_j^\mu\}_{i=1,\dots,L}$.}
\label{figure-generic_onshell_expansion}
\end{figure}

In this section, we focus exclusively on one particular asymptotic expansion: the on-shell expansion for wide-angle scattering. Specifically, we consider arbitrary graphs shown in Fig.~\ref{figure-generic_onshell_expansion}, where the external kinematics satisfy the following conditions:
\begin{eqnarray}
    &&p_i^2\sim \lambda Q^2\ \ (i=1,\dots,K),\quad q_j^2\sim Q^2\ \ (j=1,\dots,L),\quad p_{i_1}\cdot p_{i_2}\sim Q^2\ \ (i_1\neq i_2).\nonumber\\
    &&
\label{eq:wideangle_onshell_kinematics}
\end{eqnarray}
Namely, the external momenta $\{p_i^\mu\}$ have small virtualities while $\{q_j^\mu\}$ have large virtualities. The ``wide-angle'' condition $p_{i_1}\cdot p_{i_2}\sim Q^2$ further implies that $\{p_i^\mu\}$ are (nearly) aligned with distinct lightcones.

\subsection{Facet regions: an all-order result}
\label{section-facet_regions_all_order_result}

\begin{figure}[h]
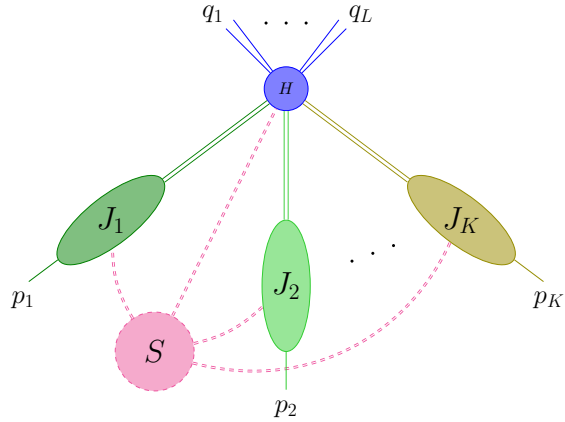

\centering
\include{figs/onshell_generic_facet_region}
\caption{The general configuration of facet regions in the on-shell expansion of wide-angle scattering, where the entire graph consists of a hard subgraph $H$, some jet subgraphs $J_1,\dots, J_K$, and a soft subgraph $S$. All the off-shell external momenta $q_1,\dots,q_L$ attach to $H$, and each on-shell external momentum $p_i$ attaches to a corresponding $J_i$.}
\label{figure-onshell_generic_facet_region}
\end{figure}

The facet region structure has been made clear at all-loop order: each loop momentum is either in the hard mode $Q(1,1,1)$, or the collinear-$i$ mode $Q(1,\lambda,\lambda^{1/2})$ (for $i=1,\dots,K$)\footnote{Note that the $Q(1,\lambda,\lambda^{1/2})$ scaling is based on aligning the lightcone direction with $p_i^\mu$.}, or the soft mode $Q(\lambda,\lambda,\lambda)$. Furthermore, the subgraphs corresponding to these modes interact according to the structure illustrated in Fig.~\ref{figure-onshell_generic_facet_region}, where each of the subgraphs $H, J_1, \dots, J_K$ must be connected, whereas $S$ may consist of multiple connected components.

Note that additional constraints of the subgraphs are needed to avoid scaleless integrals:\footnote{These requirements are first derived in Ref.~\citen{GrdHzgJnsMaSchlk22} where (1) and (2) are equivalently stated in terms of ``reduced subgraphs''.}
\begin{enumerate}
    \item [(1)] the total momentum flowing into each one-vertex irreducible (1VI) component of $H$ is off-shell;
    \item [(2)] the total momentum flowing into each 1VI component of $\widetilde{J}_i$ is of the form $(p_i+l)^\mu$, with $l^\mu$ being either zero or soft;
    \item [(3)] each connected component of $S$ is adjacent to two or more jets.
\end{enumerate}
The pattern in Fig.~\ref{figure-onshell_generic_facet_region} and the constraints above have been rigorously established as part of the ``on-shell-expansion region theorem'' in Ref.~\citen{Ma23}, which is derived through the graph-theoretical approach outlined in Section~\ref{section-identifying_facet_regions}. Below we offer several important observations regarding this theorem.

First, Fig.~\ref{figure-onshell_generic_facet_region} provides a natural representation of regions in wide-angle kinematics. As noted at the beginning of Section~\ref{section-identifying_hidden_regions}, infrared regions (those except the hard region) correspond to neighborhoods of infrared singularities characterized by Landau equations. For generic wide-angle scattering, solutions to the Landau equations align with the Coleman-Norton interpretation~\cite{ClmNtn65}, consistent with the configuration shown in Fig.~\ref{figure-onshell_generic_facet_region}.

Second, Fig.~\ref{figure-onshell_generic_facet_region} reveals a highly nontrivial result of the mode structure. In principle, describing the neighborhood of an infrared singularity can be ambiguous: it is possible to modify some jet loop momenta into alternative momentum scalings like $Q(1,\lambda^a,\lambda^b)$ with $(a,b)\neq(1,\frac{1}{2})$, or modify some soft loop momenta into $Q(\lambda^a,\lambda^b,\lambda^c)$ with $(a,b,c)\neq (1,1,1)$ while keeping momentum conservation. However, such modes always lead to scaleless integrals thus are never relevant in the on-shell expansion. If one changes the external kinematics such that some $p_i^2\sim \lambda^2 Q^2$ for some special $i$, then the mode structure can no longer be described by the hard, collinear, and soft modes above.

Third, Fig.~\ref{figure-onshell_generic_facet_region} describes the complete set of regions, which is independent of the numerator or spacetime dimension. As a result, the same list of regions applies to both scalar theory and gauge theory, with the only difference lying in the orders of the regions: those leading in one theory may be (sub)$^n$leading in the other.

For example, in Fig.~\ref{figure-onshell_generic_facet_region} we have specified some soft edges attached to the hard subgraph $H$ directly. In four-dimensional gauge theory, each such soft edge induces an additional power suppression to the Feynman integral~\cite{LbyStm78,Stm78I,Stm95book}, rendering the corresponding regions finite as $\lambda\to 0$. However, such suppression does not necessarily exist for regions in scalar theory. This also indicates a key difference between the on-shell-expansion region theorem and QCD factorization of infrared divergences: the former applies to all orders of the expansion parameter $\lambda$, while the latter applies to the leading contributions, which are logarithmically divergent as $\lambda\to 0$.

Finally, we note another key difference between the on-shell-expansion region theorem and the QCD hard-collinear-soft factorization theorem~\cite{Sen83,StmTjd-Yms03,DxnMgnStm08,FgeSwtz14,EdgStm15,Ma20}:
\begin{eqnarray}
\label{eq:hard_collinear_soft_factorization}
    \mathcal{M} = \mathcal{H}\cdot \frac{\mathcal{J}_\textup{parton}}{\mathcal{J}_\textup{eikonal}}\cdot \mathcal{S}.
\end{eqnarray}
In the equation above, $\mathcal{M}$ denotes a generic wide-angle scattering amplitude, $\mathcal{H}$ contains no infrared singularities, the ratio $\frac{\mathcal{J}_\textup{parton}}{\mathcal{J}_\textup{eikonal}}$ encodes collinear singularities, and $\mathcal{S}$ captures soft singularities. This local factorization, provable to all orders via the infrared forest formula~\cite{Ma20}, matches singularities \emph{at the integrand level}. In contrast, the on-shell expansion region theorem operates globally \emph{at the integral level} (recall that we require scaless integrals to vanish, implying that $\frac{1}{\epsilon_\textup{UV}} - \frac{1}{\epsilon_\textup{IR}} =0$), as part of the MoR.

\subsection{Hidden regions: the Landshoff scattering picture}
\label{section-hidden_regions_landshoff_scattering_picture}

We now examine the hidden regions in wide-angle scattering. For simplicity, this subsection focuses primarily on the $2 \to 2$ case, where the incoming momenta $p_1^\mu$ and $p_2^\mu$ scatter into the outgoing momenta $p_3^\mu$ and $p_4^\mu$.

As described in Section~\ref{section-identifying_hidden_regions}, an search algorithm exists to identify graphs that may exhibit pinch singularities in parameter space. At the one- and two-loop levels, we find that \emph{no} graphs exhibit such singularities; hence, hidden regions are not present. The on-shell-expansion region theorem, discussed in Section~\ref{section-facet_regions_all_order_result}, is sufficient to characterize all relevant regions.

\begin{figure}[h]
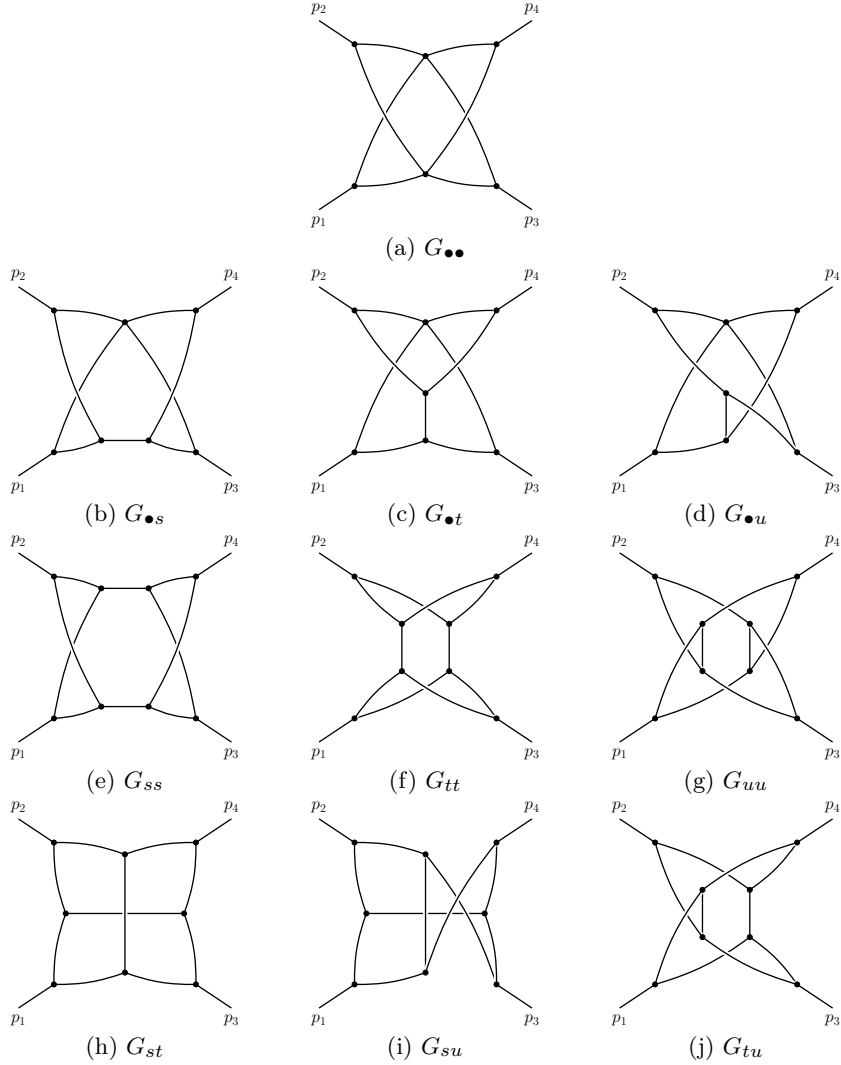

\centering
\begin{subfigure}[b]{0.25\textwidth}
\centering
\include{figs/possibly_hidden_singular_graph_dot_dot}
\vspace{-3em}\caption{$G_{\bullet \bullet}$}
\label{possibly_hidden_singular_graph_dot_dot}
\end{subfigure}
\\
\begin{subfigure}[b]{0.25\textwidth}
\centering
\include{figs/possibly_hidden_singular_graph_dot_s}
\vspace{-3em}\caption{$G_{\bullet s}$}
\label{possibly_hidden_singular_graph_dot_s}
\end{subfigure}
\qquad
\begin{subfigure}[b]{0.25\textwidth}
\centering
\include{figs/possibly_hidden_singular_graph_dot_t}
\vspace{-3em}\caption{$G_{\bullet t}$}
\label{possibly_hidden_singular_graph_dot_t}
\end{subfigure}
\qquad
\begin{subfigure}[b]{0.25\textwidth}
\centering
\include{figs/possibly_hidden_singular_graph_dot_u}
\vspace{-3em}\caption{$G_{\bullet u}$}
\label{possibly_hidden_singular_graph_dot_u}
\end{subfigure}
\\
\begin{subfigure}[b]{0.25\textwidth}
\centering
\include{figs/possibly_hidden_singular_graph_s_s}
\vspace{-3em}\caption{$G_{s s}$}
\label{possibly_hidden_singular_graph_s_s}
\end{subfigure}
\qquad
\begin{subfigure}[b]{0.25\textwidth}
\centering
\include{figs/possibly_hidden_singular_graph_t_t}
\vspace{-3em}\caption{$G_{t t}$}
\label{possibly_hidden_singular_graph_t_t}
\end{subfigure}
\qquad
\begin{subfigure}[b]{0.25\textwidth}
\centering
\include{figs/possibly_hidden_singular_graph_u_u}
\vspace{-3em}\caption{$G_{u u}$}
\label{possibly_hidden_singular_graph_u_u}
\end{subfigure}
\\
\begin{subfigure}[b]{0.25\textwidth}
\centering
\include{figs/possibly_hidden_singular_graph_s_t}
\vspace{-3em}\caption{$G_{s t}$}
\label{possibly_hidden_singular_graph_s_t}
\end{subfigure}
\qquad
\begin{subfigure}[b]{0.25\textwidth}
\centering
\include{figs/possibly_hidden_singular_graph_s_u}
\vspace{-3em}\caption{$G_{s u}$}
\label{possibly_hidden_singular_graph_s_u}
\end{subfigure}
\qquad
\begin{subfigure}[b]{0.25\textwidth}
\centering
\include{figs/possibly_hidden_singular_graph_t_u}
\vspace{-3em}\caption{$G_{t u}$}
\label{possibly_hidden_singular_graph_t_u}
\end{subfigure}
\caption{All the massless four-point three-loop graphs with a possible pinch Landau singularity in parameter space. The first figure $G_{\bullet\bullet}$ is an eight-edge graph, $G_{\bullet s}$, $G_{\bullet t}$, and $G_{\bullet u}$ are nine-edge graphs, while the remaining six are ten-edge graphs. Note that by expanding the four-point vertices of $G_{\bullet \bullet}$, one can obtain the other nine.}
\label{figure-possibly_hidden_singular_graphs}
\end{figure}

At three-loop level, there are \emph{ten graphs} with potential pinch singularities. These graphs are nonplanar, as shown in Fig.~\ref{figure-possibly_hidden_singular_graphs}, where Figs.~\ref{possibly_hidden_singular_graph_dot_s}-\ref{possibly_hidden_singular_graph_t_u} can be obtained from Fig.~\ref{possibly_hidden_singular_graph_dot_dot} by expanding one or two of its four-point vertices.

\begin{figure}[h]
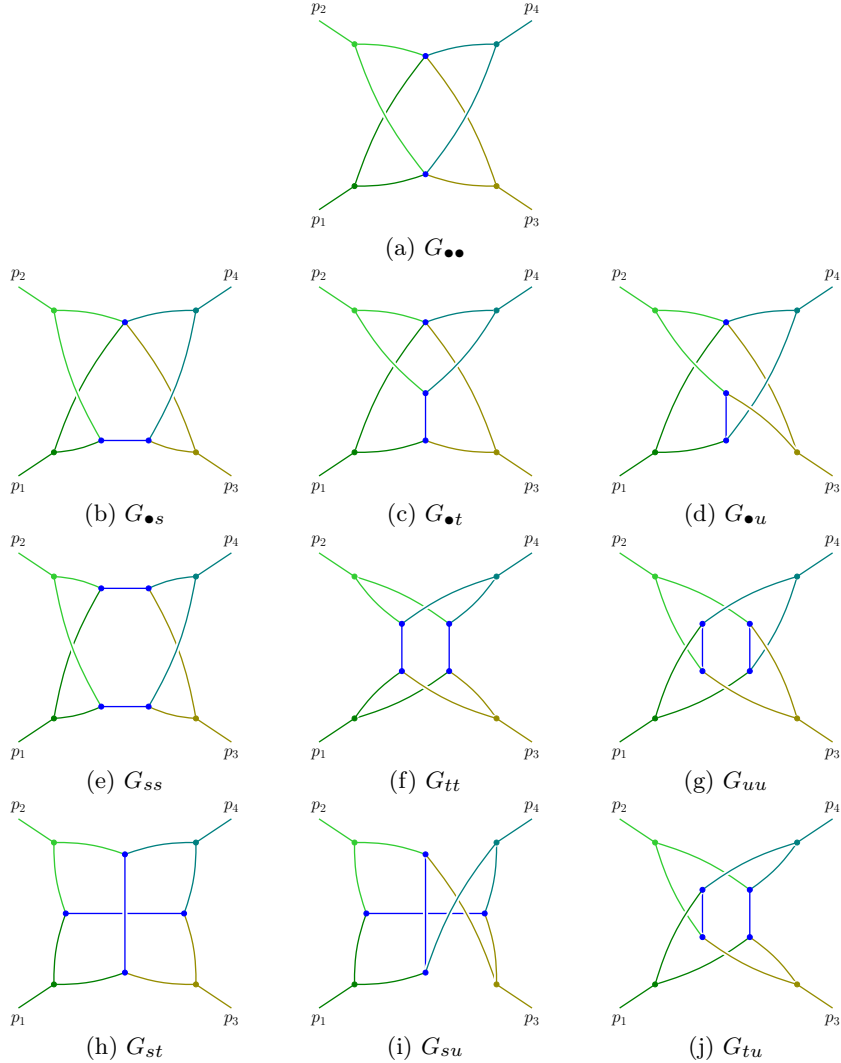

\centering
\begin{subfigure}[b]{0.25\textwidth}
\centering
\include{figs/possibly_hidden_singular_graph_dot_dot_wideangle}
\vspace{-3em}\caption{$G_{\bullet \bullet}$}
\label{possibly_hidden_singular_graph_dot_dot_wideangle}
\end{subfigure}
\\
\begin{subfigure}[b]{0.25\textwidth}
\centering
\include{figs/possibly_hidden_singular_graph_dot_s_wideangle}
\vspace{-3em}\caption{$G_{\bullet s}$}
\label{possibly_hidden_singular_graph_dot_s_wideangle}
\end{subfigure}
\qquad
\begin{subfigure}[b]{0.25\textwidth}
\centering
\include{figs/possibly_hidden_singular_graph_dot_t_wideangle}
\vspace{-3em}\caption{$G_{\bullet t}$}
\label{possibly_hidden_singular_graph_dot_t_wideangle}
\end{subfigure}
\qquad
\begin{subfigure}[b]{0.25\textwidth}
\centering
\include{figs/possibly_hidden_singular_graph_dot_u_wideangle}
\vspace{-3em}\caption{$G_{\bullet u}$}
\label{possibly_hidden_singular_graph_dot_u_wideangle}
\end{subfigure}
\\
\begin{subfigure}[b]{0.25\textwidth}
\centering
\include{figs/possibly_hidden_singular_graph_s_s_wideangle}
\vspace{-3em}\caption{$G_{s s}$}
\label{possibly_hidden_singular_graph_s_s_wideangle}
\end{subfigure}
\qquad
\begin{subfigure}[b]{0.25\textwidth}
\centering
\include{figs/possibly_hidden_singular_graph_t_t_wideangle}
\vspace{-3em}\caption{$G_{t t}$}
\label{possibly_hidden_singular_graph_t_t_wideangle}
\end{subfigure}
\qquad
\begin{subfigure}[b]{0.25\textwidth}
\centering
\include{figs/possibly_hidden_singular_graph_u_u_wideangle}
\vspace{-3em}\caption{$G_{u u}$}
\label{possibly_hidden_singular_graph_u_u_wideangle}
\end{subfigure}
\\
\begin{subfigure}[b]{0.25\textwidth}
\centering
\include{figs/possibly_hidden_singular_graph_s_t_wideangle}
\vspace{-3em}\caption{$G_{s t}$}
\label{possibly_hidden_singular_graph_s_t_wideangle}
\end{subfigure}
\qquad
\begin{subfigure}[b]{0.25\textwidth}
\centering
\include{figs/possibly_hidden_singular_graph_s_u_wideangle}
\vspace{-3em}\caption{$G_{s u}$}
\label{possibly_hidden_singular_graph_s_u_wideangle}
\end{subfigure}
\qquad
\begin{subfigure}[b]{0.25\textwidth}
\centering
\include{figs/possibly_hidden_singular_graph_t_u_wideangle}
\vspace{-3em}\caption{$G_{t u}$}
\label{possibly_hidden_singular_graph_t_u_wideangle}
\end{subfigure}
\caption{The Landshoff scattering pictures corresponding to the graphs in Fig.~\ref{figure-possibly_hidden_singular_graphs}. Here the hard subgraph is colored in blue, and the jet subgraphs are colored in different types of green.}
\label{figure-possibly_hidden_singular_graphs_wideangle}
\end{figure}

As outlined in Section~\ref{section-identifying_hidden_regions}, the hidden region(s) for each of these graphs can be identified through a proper dissection of the original polytope followed by an analysis of the lower facets of the obtained sub-polytopes. This procedure is detailed in Section 3.3 of Ref.~\citen{GrdHzgJnsMa24}, and here let us only present the results: each of the ten graphs in Fig.~\ref{figure-possibly_hidden_singular_graphs} has a unique hidden region, whose configuration is the \emph{Landshoff scattering}, where hard scatterings take place at distinct places. By coloring the hard subgraph in blue, and the jet subgraphs in (different types of) green, we describe these hidden regions in Fig.~\ref{figure-possibly_hidden_singular_graphs_wideangle}.

Let us recall one remark we made in Section~\ref{section-facet_regions_hidden_regions}: the subgraph $\gamma_\textup{off}^{}$, defined as the union of connected subgraphs exchanging off-shell momentum (either hard or Glauber), is connected for facet regions while disconnected for hidden regions. We previously saw one-loop examples in Fig.~\ref{figure-hidden_region_examples}, and now provide more nontrivial three-loop graphs that support this observation.

In gauge theory amplitudes and cross sections, Landshoff scattering is power-suppressed and does not contribute to infrared divergences~\cite{LbtdStm85,BottsStm89}. However, if we consider an individual scalar Feynman integrals, the contribution from Landshoff scattering can be important. For instance, the leading contribution to Fig.~\ref{possibly_hidden_singular_graph_dot_dot} comes exclusively from the Landshoff scattering region (Fig.~\ref{possibly_hidden_singular_graph_dot_dot_wideangle}), which scales as $\lambda^{-\frac{1}{2}-3\epsilon}$ as $\lambda\sim \frac{p_i^2}{Q^2}\to 0$ (a power divergence) while other regions contribute at most logarithmic divergences. One can derive these results from a straightforward power counting, in either momentum space or parameter space.

So far, this hidden region \emph{cannot} be identified by any existing computational tools, except via the ``search algorithm + polytope dissection'' strategy described in Section~\ref{section-identifying_hidden_regions}, due to the nontrivial structure of the $\mathcal{F}$ polynomial. Specifically, all the graphs in Fig.~\ref{figure-possibly_hidden_singular_graphs} contain $\mathcal{F}$ terms of the form
\begin{eqnarray}
    -s\cdot(x_2x_5 - x_1x_6)(x_4x_7 - x_3x_8) -t\cdot(x_2x_3 - x_1x_4)(x_6x_7 - x_5x_8).
\end{eqnarray}
These terms are super-leading in the hidden region: each variable $x$ is $\mathcal{O}(\lambda^{-1})$, while each bracketed expression above scales as $\lambda^{-3/2}$ (rather than the naively expected $\lambda^{-2}$) due to cancellations. While \texttt{Asy2} has proven most useful in many contexts, it does not detect this particular cancellation pattern, as it ``checks all pairs of variables which are part of monomials with opposite sign''~\cite{JtzSmnSmn12}, an approach that is not suited for cancellations among quadratic terms.

The number of graphs exhibiting potential pinch singularities increases with loop order. At four loops, the algorithm from Section~\ref{section-identifying_hidden_regions} identifies 1081 such graphs, all of which contain sub-topologies from Fig.~\ref{figure-possibly_hidden_singular_graphs}. In other words, \emph{all} four-loop graphs with hidden regions can be constructed by adding one loop to one of the ten three-loop base graphs. This strongly suggests that Landshoff scattering configurations \emph{exclusively} characterize hidden regions in the $2 \to 2$ wide-angle kinematics.

\begin{figure}[h]
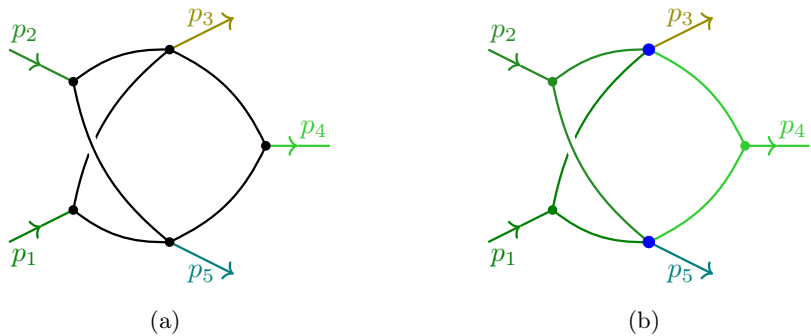

\centering
\begin{subfigure}[b]{0.4\textwidth}
\centering
\include{figs/2to3_graph}
\vspace{-1em}
\caption{}
\label{2to3_graph}
\end{subfigure}
\hspace{3em}
\begin{subfigure}[b]{0.4\textwidth}
\centering
\include{figs/2to3_Landshoff}
\vspace{-1em}
\caption{}
\label{2to3_Landshoff}
\end{subfigure}
\caption{A two-loop graph (a) describing the $2\to3$ scattering in the wide-angle kinematics and its associated Landshoff scattering region (b), where the two hard scattering vertices are colored in blue and the nontrivial jets $J_1,J_2,J_4$ are colored in different types of green.}
\label{figure-2to3_scattering_with_Landshoff}
\end{figure}

So far, we have discussed only the $2 \to 2$ scattering case. In fact, similar conclusions apply to a broader class of hard processes, for example, $2 \to 3$ wide-angle scattering. In this scenario, graphs with hidden regions begin to appear at two loops, with one example shown in Fig.~\ref{2to3_graph}. The corresponding hidden region exhibits the following pattern: each incoming momentum ($p_1^\mu, p_2^\mu$) splits into two collinear momenta, which undergo separate hard scatterings. At these distinct scattering points, respectively, the outgoing momenta $p_3^\mu$ and $p_5^\mu$ are emitted, while the remaining momenta, which are collinear to each other, eventually merge to form the outgoing momentum $p_4^\mu$. Clearly, this hidden region configuration (Fig.~\ref{2to3_Landshoff}) follows the Landshoff scattering pattern.

\begin{figure}[h]
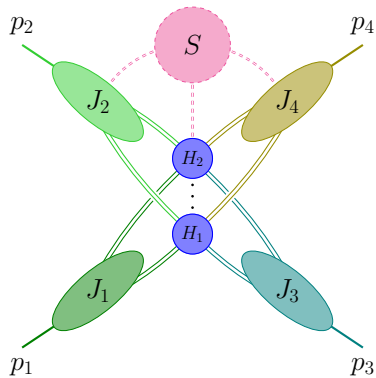

\centering
\include{figs/onshell_generic_hidden_region}
\vspace{-1cm}\caption{The general configuration of hidden regions in the on-shell expansion of $2\to 2$ wide-angle scattering. Similar to facet regions (see Fig.~\ref{figure-onshell_generic_facet_region}), the entire graph consists of a hard subgraph $H$, the jet subgraphs $J_1,\dots, J_4$, and the soft subgraph $S$. Meanwhile, here the hard subgraph has multiple connected components $H_1,H_2,\dots$, and all four jets are attached to each of these components. We note that this picture has been confirmed at the three-loop level, but remains a conjecture for higher loops.}
\label{figure-onshell_generic_hidden_region}
\end{figure}

Based on the evidence above, we conjecture that for generic wide-angle kinematics, all hidden regions exhibit a Landshoff-scattering configuration. See Fig.~\ref{figure-onshell_generic_hidden_region} for the case of $2 \to 2$ scattering. For graphs with more loops or external legs, hard scatterings can occur at three or more distinct locations, where the corresponding subgraphs would include $H_3$, $H_4$, and so on.

\section{Other asymptotic expansions in the wide-angle kinematics}
\label{section-other_asymptotic_expansions_wideangle_scenario}

In this section, we continue investigating asymptotic expansions in wide-angle kinematics. We shall focus particularly on three types of expansion: the soft expansion (where soft external momenta are included), the timelike-collinear expansion (where initial- or final-state collinear momenta are included), and the mass expansion for heavy-to-light decay processes (where internal masses are included). Compared to the on-shell expansion, these scenarios reveal richer and more nuanced region structures, reflecting the additional subgraph constraints by external kinematics, as imposed by each expansion type.

\subsection{The soft expansion}
\label{section-soft_expansion}

\begin{figure}[t]
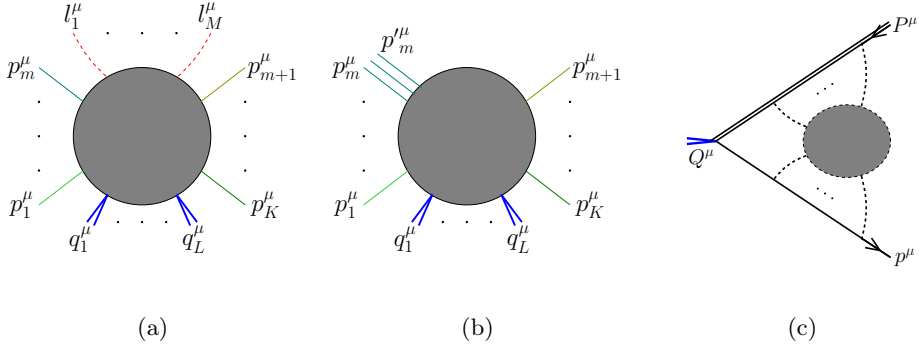

\centering
\begin{subfigure}[b]{0.32\textwidth}
\centering
\include{figs/generic_soft_expansion}
\vspace{-1em}
\caption{}
\label{figure-generic_soft_expansion}
\end{subfigure}
\hfill
\begin{subfigure}[b]{0.32\textwidth}
\centering
\include{figs/generic_timelike_collinear_expansion}
\vspace{-1em}
\caption{}
\label{figure-generic_timelike_collinear_expansion}
\end{subfigure}
\hfill
\begin{subfigure}[b]{0.32\textwidth}
\centering
\include{figs/generic_mass_expansion}
\vspace{-1.5em}
\caption{}
\label{figure-generic_mass_expansion}
\end{subfigure}
\caption{The graphs representing three other asymptotic expansions in the wide-angle kinematics, which we focus on in Section~\ref{section-other_asymptotic_expansions_wideangle_scenario}: (a) the soft expansion, (b) the timelike-collinear expansion, and (c) the mass expansions (for heavy-to-light decay). The gray blob in each figure contains an arbitrary number of internal edges and vertices.}
\label{figure-other_expansions_wideangle_kinematics}
\end{figure}

Here we consider Feynman graphs with massless propagators and $K+L+M$ external momenta, denoted as $p_1^\mu,\dots,p_K^\mu, q_1^\mu,\dots,q_L^\mu, l_1^\mu,\dots,l_M^\mu$ (see Fig.~\ref{figure-generic_soft_expansion}). Slightly different from the on-shell expansion kinematics, each $q_j^\mu$ is off-shell, and each $p_i^\mu$ and $l_k^\mu$ is strictly on-shell ($p_i^2 = l_k^2 = 0$). The key distinction between $p_i^\mu$ and $l_k^\mu$ lies in the fact that each $p_i^\mu$ is collinear to one specific direction, whereas each $l_k^\mu$ is soft. Therefore, the kinematics here can be obtained from the on-shell expansion kinematics by setting all the $p_i^2$ to zero and including soft external momenta. Some recent advances in soft current calculations, building on this soft expansion framework, include Refs.~\citen{Zhu20,CtnCieriCfrCrdsch23,CzkEschmSchlbg23,CtnCfrTrn20,DDcDuhrHndLiu23,HzgMaMstbgSrsh23,ChenLuoYangZhu24,ChenLiu24}.

The soft expansion is then defined as the limit where all the components of the soft external momenta vanish at the same speed. Equivalently, by introducing the small parameter $\lambda\ll 1$,
\begin{subequations}
\label{eq:wideangle_soft_kinematics}
\begin{align}
    & p_i^2=0\ \ (i=1,\dots,K), \quad q_j^2\sim Q^2\ \ (j=1,\dots,L), \quad l_k^2=0\ \ (k=1,\dots,M),\\
    & p_{i_1}\cdot p_{i_2}\sim Q^2\ \ (i_1\neq i_2), \quad p_i\cdot l_k\sim q_j\cdot l_k\sim \lambda Q^2, \quad l_{k_1}\cdot l_{k_2}\sim \lambda^2 Q^2\ \ (k_1\neq k_2).
\end{align}
\end{subequations}

We first note that the soft expansion can be seen as a generalization of the on-shell expansion. This relationship is due to the defining properties of their external kinematics. In the soft expansion, each $p_i$ is \emph{exactly on shell}, i.e., $(p_i^\text{soft})^2=0$; furthermore, we have $(p_i^\text{soft}+l_k)^2 = 2p_i^\text{soft}\cdot l_k \sim \lambda Q^2$ for a soft momentum $l_k$. In contrast, in the on-shell expansion, each $p_i$ is \emph{close to a lightcone} instead, i.e., $(p_i^\text{onshell})^2\sim \lambda Q^2\neq 0$, thus such $p_i^\text{onshell}$ can be seen as $p_i^\text{soft}+l_k$. Therefore, the kinematics of the on-shell expansion is equivalent to the kinematics of the soft expansion regarding a special set of graphs, where the on-shell external momenta $\{p_i\}$ and soft external momenta $\{l_i\}$ form pairs, each entering the same vertex of~$G$.

The soft expansion can be seen as a generalization of the on-shell expansion. This relationship is due to their definitions, (\ref{eq:wideangle_onshell_kinematics}) and (\ref{eq:wideangle_soft_kinematics}). In the soft expansion, each $p_i$ is \emph{exactly on shell}, i.e., $(p_i^\text{soft})^2=0$; furthermore, we have $(p_i^\text{soft}+l_k)^2 = 2p_i^\text{soft}\cdot l_k \sim \lambda Q^2$ for a soft momentum $l_k$. In contrast, in the on-shell expansion, each $p_i$ is \emph{close to a lightcone} instead, i.e., $(p_i^\text{onshell})^2\sim \lambda Q^2\neq 0$, thus such $p_i^\text{onshell}$ can be seen as $p_i^\text{soft}+l_k$. Therefore, the kinematics of the on-shell expansion is equivalent to the kinematics of the soft expansion regarding a special set of graphs, where the on-shell external momenta $\{p_i\}$ and soft external momenta $\{l_i\}$ form pairs, each entering the same vertex of $G$.

\begin{figure}[h]
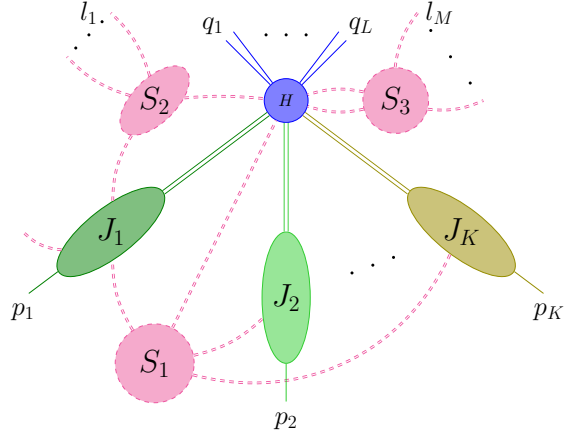

\centering
\include{figs/soft_generic_facet_region}
\caption{The general configuration of facet regions in the soft expansion of wide-angle scattering, where the entire graph consists of a hard subgraph $H$, some jet subgraphs $J_1,\dots, J_K$, and a (possibly disconnected) soft subgraph $S$. All the off-shell external momenta $q_1,\dots,q_L$ attach to $H$, each lightlike external momentum $p_i$ attaches to a corresponding $J_i$, and the soft momenta $l_1,\dots,l_M$ can attach to any vertex of $G$.  In contrast to the facet regions in the on-shell expansion (see Fig.~\ref{figure-onshell_generic_facet_region}), some soft components may be adjacent to only one, or even zero jet subgraph.}
\label{figure-soft_generic_facet_region}
\end{figure}

The facet region structure of the soft expansion can be described by Fig.~\ref{figure-soft_generic_facet_region}, as proved in Ref.~\citen{Ma23}. As a key difference from the on-shell expansion, here each soft component can attach to only the hard subgraph, or only one jet subgraph. More precisely, any soft component attached by zero or one soft external momentum must be adjacent to two or more jets; any soft component attached by two or more soft external momenta does not have this constraint.

\begin{figure}[h]
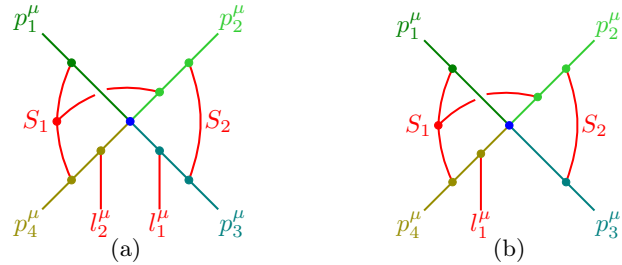

\centering
\begin{subfigure}[b]{0.3\textwidth}
\centering
\include{figs/soft_compatible_example}
\vspace{-2em}
\caption{}
\label{soft_compatible_example}
\end{subfigure}
\hspace{3em}
\begin{subfigure}[b]{0.3\textwidth}
\centering
\include{figs/soft_compatible_counter_example}
\vspace{-2em}
\caption{}
\label{soft_compatible_counter_example}
\end{subfigure}
\caption{Examples to demonstrate the requirement of the jets, where (a) shows a configuration where each jet does ``receive'' the external soft information, while (b) shows a configuration where $J_3$ does not receive any external soft information. Thus (a) is a valid region while (b) is not.}
\label{figure-soft_compatible_example_counterexample}
\end{figure}

There are some additional requirements of the jet subgraph: each jet must be both ``regular'' and ``soft compatible'', as formally defined in Ref.~\citen{Ma23}. Rather than presenting rigorous definitions here, we illustrate these concepts metaphorically: each jet must ``receive'' information from soft external momenta, and any soft component connecting three or more jets may serve as a ``messenger''. This can be demonstrated by the very similar examples shown in Fig.~\ref{figure-soft_compatible_example_counterexample}. Note that for both \ref{soft_compatible_example} and \ref{soft_compatible_counter_example}, the external momenta $l_1^\mu$ and $l_2^\mu$ are soft, and there are two soft components, $S_1$ (adjacent to three jets $J_1,J_2,J_4$) and $S_2$ (adjacent to two jets $J_2,J_3$). From the criteria above, $S_1$ qualifies as a messenger while $S_2$ does not. In \ref{soft_compatible_example}, the jets $J_3$ and $J_4$ receive information from soft external momenta directly, and $S_1$ delivers the information from $J_4$ to $J_1$ and $J_2$. In contrast, despite having identical messenger behavior through $S_1$ in \ref{soft_compatible_counter_example}, $J_3$ ultimately receives no soft information. Consequently, \ref{soft_compatible_example} is a valid region while \ref{soft_compatible_counter_example} is not: expansions based on \ref{soft_compatible_counter_example} would contain a scaleless integral.

We emphasize that this jet subgraph requirement applies broadly to Feynman integrals beyond QCD. For instance, analogous constraints on non-vanishing loop corrections have been observed in the study of Soft-Collinear Gravity~\cite{BnkHgrSzfr22gravitational}. This universality is expected, as we have remarked in Section~\ref{section-facet_regions_all_order_result}, the complete set of regions depends solely on the singularity structure of the integrand, and is independent of both numerator and spacetime dimension.

\subsection{The timelike-collinear expansion}
\label{section-timelike_collinear_expansion}

The region analysis can be extended to kinematics involving \emph{timelike collinear} momenta, which is defined as configurations where incoming (or outgoing) external momenta are mutually collinear. This differs from \emph{spacelike collinear} kinematics, where incoming momenta become collinear with outgoing ones, a case with more subtleties which we will explain in Section~\ref{section-spacelike_collinear_kinematics_Glauber_singularities}.

In general cases, the setup may include off-shell external momenta $q_1^\mu,\dots,q_L^\mu$, on-shell external momenta $p_1^\mu,\dots,p_K^\mu$ pointing in different directions, and additional on-shell external momenta $\{ {p'}_k^\mu \}$, each of which collinear to one of $\{p_1^\mu,\dots,p_K^\mu\}$ (see Fig.~\ref{figure-generic_timelike_collinear_expansion} for an example). The corresponding kinematics are:
\begin{subequations}
\label{eq:wideangle_collinear_kinematics}
\begin{align}
    & p_i^2=0={p'}_k^2\ \ (\forall i,k), \quad q_j^2\sim Q^2\ \ (\forall j),\quad p_{i_1}\cdot p_{i_2}\sim Q^2\ \ (\forall i_1\neq i_2),\\
    & p_i\cdot {p'}_k\sim \lambda Q^2\ \textup{ for }{p'}_k \parallel p_i^\mu,\qquad p_i\cdot {p'}_k\sim Q^2\ \textup{ for }{p'}_k\nparallel p_i^\mu.
\end{align}
\end{subequations}
This asymptotic expansion in the small parameter $\lambda$ defines the (timelike-)collinear expansion~\cite{EbtMstbgVita20collinear,EbtMstbgVita20transverse,EbtMstbgVita20Njettiness,EbtMstbgVita21TMD,GuanHzgMaMstbgSrsh25}. The general region configuration resembles Fig.~\ref{figure-soft_generic_facet_region}, but with soft external momenta replaced by lightlike momenta that may be mutually timelike collinear. Analogous to the soft-expansion case, an additional requirement of the jet applies to avoid scaleless integrals: each jet must be both regular and collinear-compatible, as formally defined in Ref.~\citen{Ma23}.

The similarity between regions in the soft expansion and those in the timelike-collinear expansion arises from their shared origin of $\mathcal{O}(\lambda)$ energy scales. For the soft expansion, $(p+l)^2\sim \lambda$, while for the timelike-collinear expansion, $(p_i+p_k')^2\sim \lambda$. In order that the jet and soft loop integrations are all scaleful, they must receive the input from some of these $\mathcal{O}(\lambda)$ scales, either directly from the external kinematics or mediated through those messenger soft subgraphs.

\subsection{The mass expansion for heavy-to-light decays}
\label{section-mass_expansion}

Our analysis has thus far focused on asymptotic expansions involving massless propagators. We now extend this framework to include massive propagators, specifically examining heavy-to-light decay processes where particle masses exhibit hierarchical scales, with the expansion parameter $\lambda$ defined as their ratio.

The relevant graphs for this scenario are characterized in Fig.~\ref{figure-generic_mass_expansion}. There is a path connecting $Q^\mu$ and $P^\mu$, represented by the double solid line and composed exclusively of mass-$M$ edges; there is a path connecting $Q^\mu$ and $p^\mu$, represented by the single solid line and composed exclusively of mass-$m$ edges; meanwhile, all the other edges of $G$, represented by the dashed lines, are massless. By introducing the small parameter $\lambda\sim m^2/M^2 \ll 1$, we have
\begin{eqnarray}
    P^2=M^2\sim Q^2,\quad p^2=m^2\sim \lambda Q^2,\quad P\cdot p\sim Q^2.
\label{eq:decay_mass_kinematics}
\end{eqnarray}

The region structure in the mass expansion (\ref{eq:decay_mass_kinematics}) exhibits significantly greater complexity in its momentum modes. Unlike the simpler hard-collinear-soft structure of previous asymptotic expansions, this case features an unbounded proliferation of modes as the loop order increases, a phenomenon sometimes referred to as the \emph{cascade of modes}~\cite{FtsSfr24,JskwJnsSzfUrch25}. This behavior originates from the exact on-shell condition of the two external massive particles. Table~\ref{table-mode_structure_mass_expansion} enumerates some representative modes and their emergence loop orders.
\begin{table}[h]
\tbl{Momentum modes involved in the mass expansion (\ref{eq:decay_mass_kinematics}) and their corresponding starting loops.\label{table-mode_structure_mass_expansion}}
{\begin{tabular}{@{}l l@{}} \toprule
{\bf Momentum mode} & {\bf Starting from} \\ \colrule
hard: $Q(1,1,1)$ & 1 loop \\
collinear: $Q(1,\lambda,\lambda^{1/2})$ & 1 loop \\
soft: $Q(\lambda,\lambda,\lambda)$ & 2 loops \\
semihard: $Q(\lambda^{1/2},\lambda^{1/2},\lambda^{1/2})$ & 2 loops \\
soft$\cdot$collinear: $Q(\lambda,\lambda^2,\lambda^{3/2})$ & 3 loops \\
semihard$\cdot$collinear: $Q(\lambda^{1/2},\lambda^{3/2},\lambda)$ & 3 loops \\
semicollinear: $Q(1,\lambda^{1/2},\lambda^{1/4})$ & 3 loops \\
soft$^2$: $Q(\lambda^2,\lambda^2,\lambda^2)$ & 4 loops \\ 
$\dots$ & $\dots$ \\ \botrule
\end{tabular}}
\end{table}

\begin{figure}[h]
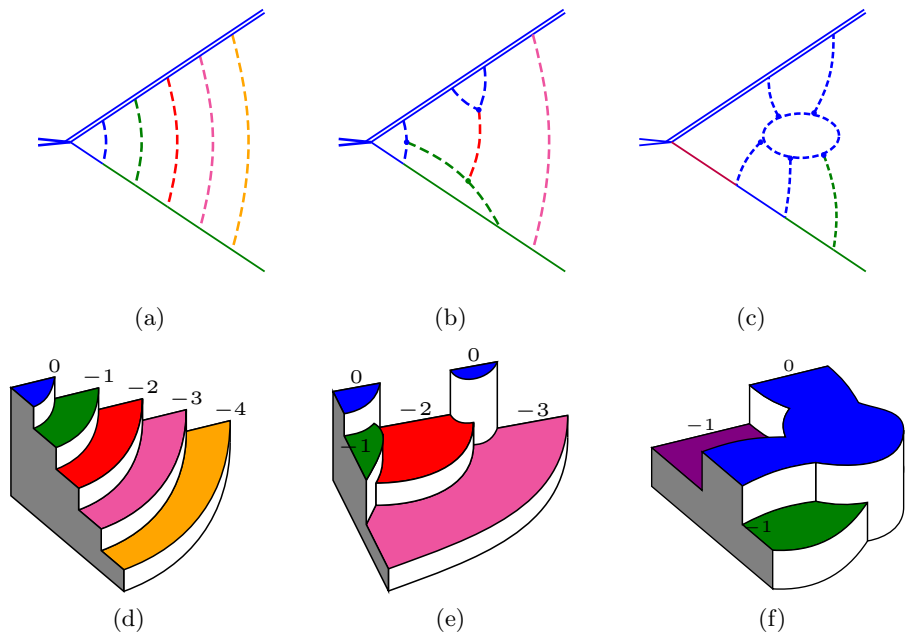

\centering
\begin{subfigure}[b]{0.25\textwidth}
\centering
\include{figs/mass_expansion_example_5loop_graph1}
\vspace{-2em}\caption{}
\label{mass_expansion_example_5loop_graph1}
\end{subfigure}
\qquad
\begin{subfigure}[b]{0.25\textwidth}
\centering
\include{figs/mass_expansion_example_5loop_graph2}
\vspace{-2em}\caption{}
\label{mass_expansion_example_5loop_graph2}
\end{subfigure}
\qquad
\begin{subfigure}[b]{0.25\textwidth}
\centering
\include{figs/mass_expansion_example_5loop_graph3}
\vspace{-2em}\caption{}
\label{mass_expansion_example_5loop_graph3}
\end{subfigure}
\\
\begin{subfigure}[b]{0.3\textwidth}
\include{figs/mass_expansion_example_5loop_graph1_terrace}
\vspace{-1em}\caption{}
\label{figure-mass_expansion_example_5loop_graph1_terrace}
\end{subfigure}
\quad
\begin{subfigure}[b]{0.3\textwidth}
\include{figs/mass_expansion_example_5loop_graph2_terrace}\vspace{10pt}
\vspace{-2em}\caption{}
\label{figure-mass_expansion_example_5loop_graph2_terrace}
\end{subfigure}
\quad
\begin{subfigure}[b]{0.3\textwidth}
\include{figs/mass_expansion_example_5loop_graph3_terrace}\vspace{10pt}
\vspace{-2em}\caption{}
\label{figure-mass_expansion_example_5loop_graph3_terrace}
\end{subfigure}
\caption{An illustration of the terrace formalism. Figs.~(a), (b), and (c) depict three examples of regions in the mass expansion of five-loop graphs, where we have colored each hard-mode edge in \textbf{\color{Blue}blue}, each collinear-mode edge in \textbf{\color{Green}green}, each soft-mode edge in \textbf{\color{Red}red}, each collinear$\cdot$soft-mode edge in \textbf{\color{Rhodamine}rhodamine}, each soft${}^2$-mode edge in \textbf{\color{Orange}orange}, and each semihard-mode edge in \textbf{\color{Purple}purple}. Figs.~(d), (e), and (f) show their corresponding terraces, with surfaces colored according to the associated modes of the loop momenta.}
\label{figure-mass_expansion_example_5loop_graphs}
\end{figure}

The complex mode structure makes generic regions in the mass expansion particularly intricate. While no all-order analysis of these regions currently exists, an elegant \emph{terrace formalism} provides a useful characterization for \emph{planar} Feynman graphs. This approach visualizes momentum modes as ``descending steps'': starting from loops farthest from the external momenta $P^\mu$ and $p^\mu$ and approaching them, each successive loop exhibits progressively softer momentum scaling, analogous to descending a staircase. Consequently, any two-dimensional planar graph can be represented as a three-dimensional terrace structure (see Fig.~\ref{figure-mass_expansion_example_5loop_graphs} for representative examples), where subgraph requirements for regions correspond to specific surface conditions on each terrace.

\section{Regions in the Regge-limit expansion}
\label{section-regions_Regge_limit_expansion}

Having explored various region structures for asymptotic expansions in wide-angle kinematics, we now turn to spacelike collinear kinematics, where external momenta are aligned along lightlike directions with spacelike separations (initial collinear to final), with a particular focus on $2\to2$ forward scattering. Section~\ref{section-spacelike_collinear_kinematics_Glauber_singularities} lays out some foundational aspects of Glauber singularities in generic $m\to n$ scattering with spacelike collinear momenta. Section~\ref{section-2to2_forward_scattering} then specializes to the region structure of $2\to2$ forward scattering, addressing both facet and hidden regions.

\subsection{Spacelike collinear kinematics and Glauber singularities}
\label{section-spacelike_collinear_kinematics_Glauber_singularities}

Glauber singularities arise from momentum configurations typically characterized by anisotropic scaling $k^\mu\sim (\lambda,\lambda,\lambda^{1/2})$ in lightcone coordinates, where longitudinal components are $\mathcal{O}(\lambda^{1/2})$ while transverse components are $\mathcal{O}(\lambda)$. These singularities differ fundamentally from soft singularities $(\lambda,\lambda,\lambda)$ that exhibit isotropic scaling. As a crucial property, Glauber singularities are usually responsible for the violation of collinear factorization~\cite{CtnDFlRrg11,FshSmrSmk12,Schwartz:2017nmr}, at both the amplitude and cross-section level.

In generic wide-angle kinematics, Glauber singularities are absent because the integration contour can always be deformed to avoid pinches in the Glauber mode $(\lambda,\lambda,\lambda^{1/2})$~\cite{ClsStm81}. This explains why our previous region analyses in Sections~\ref{section-regions_onshell_expansion_wideangle_kinematics} and \ref{section-other_asymptotic_expansions_wideangle_scenario} required no Glauber-mode treatment.

The situation changes for spacelike collinear kinematics, where contour deformation may fail. Consider the previous example in Fig.~\ref{figure-hidden_region_example1} with $p_1^\mu$ along the $+$ direction and $p_2^\mu$ along the $-$ direction: propagators 2 and 3 pinch the $k^-$ contour at $\mathcal{O}(\lambda)$, while propagators 4 and 5 pinch the $k^+$ contour at $\mathcal{O}(\lambda)$. These unavoidable pinch surfaces necessitate the inclusion of Glauber regions in the MoR analysis.

We note that not all graphs with spacelike collinear external momenta develop Glauber singularities. For the region analysis of the $2\to2$ forward scattering, as we will see below, Glauber regions appear only from three loops (unless special topologies exist where both spacelike collinear momenta enter the same vertex). Remarkably, those three-loop graphs with hidden Glauber regions share identical topologies with those in the wide-angle kinematics, suggesting a deep correspondence between Glauber singularities and Landshoff-scattering singularities.

\subsection{The $2\to 2$ forward scattering}
\label{section-2to2_forward_scattering}

We start by defining the Regge-limit expansion in $p_1+p_2\to p_3+p_4$ forward scattering, where every external momentum is strictly massless and on shell, with $p_3$ (nearly) collinear to $p_1$ and $p_4$ nearly collinear to $p_2$, namely,
\begin{eqnarray}
\label{Regge_limit}
    p_i^2=0\ \ (i=1,2,3,4),\qquad   \left | \frac{t}{s} \right | \sim  \left | \frac{t}{u} \right | \sim \lambda,
\end{eqnarray}
Below we discuss the facet regions and hidden regions separately. For convenience, we choose the lightcone coordinate where $p_1^\mu$ and $p_3^\mu$ to be along the $+$ direction while $p_2^\mu$ and $p_4^\mu$ along the $-$ direction.

\subsubsection{Facet regions}

The facet region structure in the Regge-limit expansion also feature a cascade of modes, similar to the mass expansion case in Section~\ref{section-mass_expansion}. At one- and two-loop level, regions involve hard ($k_H^\mu\sim Q(1,1,1)$), collinear-13 ($k_{C_{13}}^\mu\sim Q(1,\lambda,\lambda^{1/2})$), and collinear-24 ($k_{C_{24}}^\mu\sim Q(\lambda,1,\lambda^{1/2})$) modes only. Note that in the special configuration where $p_1^\mu$ and $p_3^\mu$, or $p_2^\mu$ and $p_4^\mu$, attach to the same vertex, Glauber-mode momentum $k_G^\mu\sim Q(\lambda,\lambda,\lambda^{1/2})$ can arise, which is endowed by the combination of the external momenta $(p_1+p_3)^\mu=(p_2+p_4)^\mu$.

\begin{figure}[h]
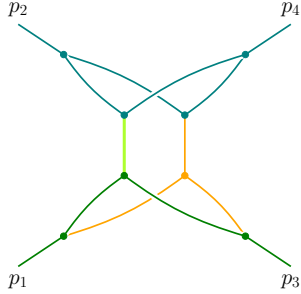

\centering
\include{figs/3loop_CS_mode_example}
\vspace{-1cm}\caption{One of the facet regions of $G_{tt}$ (notation given in Fig.~\ref{figure-possibly_hidden_singular_graphs}), which corresponds to the region vector $(-2,-3,-1,-1,-2,-3,-1,-1,-1,-3;1)$. In addition to the {\color{teal} \bf collinear-24} mode, this region also features a {\color{GreenYellow} \bf Glauber} mode, a {\color{Orange} \bf soft$\cdot$collinear-1} mode, and a {\color{Green} \bf (collinear-1)$^2$} mode, as colored in the figure.}
\label{figure-3loop_CS_mode_example}
\end{figure}

At three- and four-loop level, the following additional modes arise in loop momenta:
\begin{align}
\label{ReggeMods_3and4_loops}
\begin{cases}
\text{soft}: & k_S^\mu\sim Q(\lambda,\lambda,\lambda);\\
\text{soft}\cdot \text{collinear-1}:  \hspace*{30pt} 
& 
k_{SC_1}^\mu\sim Q(\lambda,\lambda^2,\lambda^{3/2});
\\
\text{soft}\cdot \text{collinear-2}: 
& 
k_{SC_2}^\mu\sim Q(\lambda^2,\lambda,\lambda^{3/2});
\\
\text{(collinear-1)}^2:  & 
k_{C_1C_1}^\mu\sim Q(1,\lambda^2,\lambda);
\\
\text{(collinear-2)}^2:   & 
k_{C_2C_2}^\mu\sim Q(\lambda^2,1,\lambda),
\end{cases}
\end{align}
An example is shown in Fig.~\ref{figure-3loop_CS_mode_example}, where the Glauber, collinear-24, soft$\cdot$collinear-1, and (collinear-1)$^2$ modes are involved in the line momenta. The upper jet (attached by $p_2^\mu$ and $p_4^\mu$) contains a loop, while the lower jet (attached by $p_1^\mu$ and $p_3^\mu$) is a tree graph. Propagators mediating between these jets exhibit Glauber and soft$\cdot$collinear-1 momentum scaling.

For higher loop orders, additional modes arise following a similar manner. For five- and six-loop graphs, loop momenta can also be in one of the following modes:
\begin{align}
\label{ReggeMods_5and6_loops}
\begin{cases}
\text{soft}^2\cdot \text{collinear-1}:  \hspace*{30pt} 
& 
k_{S^2C_1}^\mu\sim Q(\lambda^2,\lambda^3,\lambda^{5/2});
\\
\text{soft}^2\cdot \text{collinear-2}: 
& 
k_{S^2C_2}^\mu\sim Q(\lambda^3,\lambda^2,\lambda^{5/2});
\\
\text{(collinear-1)}^3:  & 
k_{C_1C_1}^\mu\sim Q(1,\lambda^3,\lambda^{3/2});
\\
\text{(collinear-2)}^3:   & 
k_{C_2C_2}^\mu\sim Q(\lambda^3,1,\lambda^{3/2}),
\end{cases}
\end{align}
This pattern naturally suggests that new momentum modes emerge ``incrementally'', with additional modes appearing every two-loop order.

\subsubsection{Hidden regions}

\begin{figure}[h]
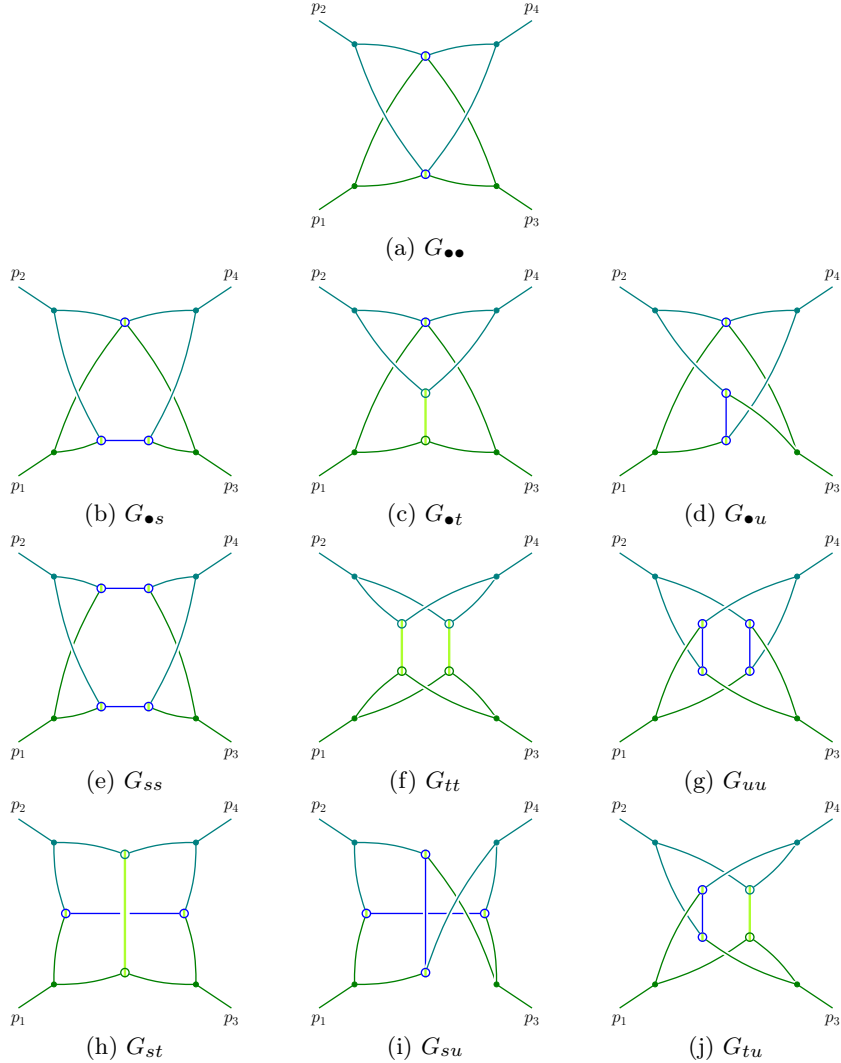

\centering
\begin{subfigure}[b]{0.25\textwidth}
\centering
\include{figs/possibly_hidden_singular_graph_dot_dot_forward}
\vspace{-3em}\caption{$G_{\bullet \bullet}$}
\label{possibly_hidden_singular_graph_dot_dot_forward}
\end{subfigure}
\\
\begin{subfigure}[b]{0.25\textwidth}
\centering
\include{figs/possibly_hidden_singular_graph_dot_s_forward}
\vspace{-3em}\caption{$G_{\bullet s}$}
\label{possibly_hidden_singular_graph_dot_s_forward}
\end{subfigure}
\qquad
\begin{subfigure}[b]{0.25\textwidth}
\centering
\include{figs/possibly_hidden_singular_graph_dot_t_forward}
\vspace{-3em}\caption{$G_{\bullet t}$}
\label{possibly_hidden_singular_graph_dot_t_forward}
\end{subfigure}
\qquad
\begin{subfigure}[b]{0.25\textwidth}
\centering
\include{figs/possibly_hidden_singular_graph_dot_u_forward}
\vspace{-3em}\caption{$G_{\bullet u}$}
\label{possibly_hidden_singular_graph_dot_u_forward}
\end{subfigure}
\\
\begin{subfigure}[b]{0.25\textwidth}
\centering
\include{figs/possibly_hidden_singular_graph_s_s_forward}
\vspace{-3em}\caption{$G_{s s}$}
\label{possibly_hidden_singular_graph_s_s_forward}
\end{subfigure}
\qquad
\begin{subfigure}[b]{0.25\textwidth}
\centering
\include{figs/possibly_hidden_singular_graph_t_t_forward}
\vspace{-3em}\caption{$G_{t t}$}
\label{possibly_hidden_singular_graph_t_t_forward}
\end{subfigure}
\qquad
\begin{subfigure}[b]{0.25\textwidth}
\centering
\include{figs/possibly_hidden_singular_graph_u_u_forward}
\vspace{-3em}\caption{$G_{u u}$}
\label{possibly_hidden_singular_graph_u_u_forward}
\end{subfigure}
\\
\begin{subfigure}[b]{0.25\textwidth}
\centering
\include{figs/possibly_hidden_singular_graph_s_t_forward}
\vspace{-3em}\caption{$G_{s t}$}
\label{possibly_hidden_singular_graph_s_t_forward}
\end{subfigure}
\qquad
\begin{subfigure}[b]{0.25\textwidth}
\centering
\include{figs/possibly_hidden_singular_graph_s_u_forward}
\vspace{-3em}\caption{$G_{s u}$}
\label{possibly_hidden_singular_graph_s_u_forward}
\end{subfigure}
\qquad
\begin{subfigure}[b]{0.25\textwidth}
\centering
\include{figs/possibly_hidden_singular_graph_t_u_forward}
\vspace{-3em}\caption{$G_{t u}$}
\label{possibly_hidden_singular_graph_t_u_forward}
\end{subfigure}
\caption{The hidden regions corresponding to the graphs in Fig.~\ref{figure-possibly_hidden_singular_graphs}, each featuring the exchange of a Glauber loop momentum between the upper and lower jets.}
\label{figure-possibly_hidden_singular_graphs_forward}
\end{figure}

To identify the graphs with hidden regions, one can apply the algorithm explained in Section~\ref{section-identifying_hidden_regions}. The result is the following. At one- and two-loop level, no graphs have pinch singularities in parameter space, thus no hidden regions can be involved. At three-loop level, there are ten graphs with potential pinch singularities, which are exactly the same ten shown in Fig.~\ref{figure-possibly_hidden_singular_graphs}. Similar to the wide-angle scattering case, each of these graphs has exactly one hidden region, as depicted in Fig.~\ref{figure-possibly_hidden_singular_graphs_forward}.

In each of these hidden regions, there is an upper jet (associated with external momenta $p_2^\mu$ and $p_4^\mu$) and a lower jet (with $p_1^\mu$ and $p_3^\mu$) exchanging momentum in the Glauber mode $Q(\lambda,\lambda,\lambda^{1/2})$. In the figures, we use circled vertices labeled \GlauberTextHard, \GlauberTextJetA, and \GlauberTextJetB\ to denote those hard, lower-jet, and upper-jet vertices respectively, through which a Glauber momentum flows in the $t$-channel. Notably, although one loop momentum is always in the Glauber mode in these graphs, it is not necessary for any line momenta to explicitly exhibit Glauber scaling. Another observation is that the subgraph $\gamma_\textup{off}^{}$ is always disconnected, as a characteristic of hidden regions we previously noted in Section~\ref{section-facet_regions_hidden_regions}.

\section{Summary and outlook}
\label{section-summary_outlook}

This paper has reviewed the systematic identification of regions in the method of regions (MoR) for asymptotic expansions of Feynman integrals, emphasizing recent advances in geometric and graph-theoretical frameworks. A central theme is the classification of regions into facet and hidden types, distinguished by their geometric origins in the Newton polytope of the Lee-Pomeransky polynomial. Facet regions, corresponding to lower facets of the polytope, are systematically identified through minimum spanning (2-)tree analyses and kinematic scaling hierarchies, while hidden regions, residing in the polytope interior, demand refined strategies involving Landau equation analyses and polytope dissections.

The all-loop-order characterization of facet regions in the on-shell expansion demonstrates the universality of the hard-collinear-soft mode structure, consistent with infrared singularity configurations. Hidden regions in this scenario appear as the Landshoff scattering, revealing an intricate connection between disconnected subgraph topologies and singularities in the parameter space arising from cancellations in the Landau equations.

Extensions to other similar expansions in the wide-angle limit underscore the sensitivity of region structures to external kinematics. For the soft and timelike-collinear expansions, facet regions require additional constraints on jet subgraphs (for example, the soft compatibility requirement of the jets) to avoid scaleless integrals. The mass expansion of the heavy-to-light decay process exhibits a cascade of momentum modes beyond the hard-collinear-soft type (e.g., semihard, soft-collinear, and semicollinear), challenging conventional factorization paradigms and practical calculations due to the proliferation of scaling hierarchies.

In the Regge-limit expansion, where external kinematics involve spacelike-collinear momenta, facet regions also exhibit a cascade of modes (e.g., soft, soft$\cdot$collinear, and collinear$^2$), while hidden regions feature Glauber loop momenta exchanging between two jets. These hidden Glauber regions, due to unavoidable Glauber singularities in spacelike kinematics, occur from three loops, and particularly, mirror the Landshoff scattering (hidden) regions observed in the wide-angle kinematics.

The expansions discussed in this review represent only a subset of the broader landscape of asymptotic expansions. We hope that the geometric and combinatorial strategies described here will serve as a foundation for analyzing other expansions in a similarly systematic fashion. On the computational side, packages implementing the method of regions, such as \texttt{AMFlow}\cite{LiuMa23AMFlow}, \texttt{DiffExp}\cite{Hdg21DiffExp}, \texttt{SeaSyde}~\cite{AmdlBcnDvtRanaVcn23SeaSyde}, \texttt{AsyInt}~\cite{Zhang24AsyInt}, etc., may benefit from incorporating these general prescriptions for region identification in specific kinematic limits. Such integrations could improve the robustness and efficiency of automated expansion techniques and deepen the interplay between analytic insights and numerical implementations.

\bigbreak
Based on the above understanding of region structures, several directions merit further exploration. Below let me list a few representatives to the best of my current knowledge.

Once the region structure for a given asymptotic expansion is known, one can develop a purely graph-finding algorithm to enumerate all regions, circumventing the need of constructing polytopes. Such an algorithm can significantly improve efficiency and has already been developed for the on-shell expansion, which is implemented in Maple and available from a public Git repository~\cite{git-Maple_file}. Ongoing work is extending this algorithm to other expansions and processes.

An all-order region analysis provides crucial insights into various aspects of the QCD infrared structure, which we shall not enumerate here. Meanwhile, the momentum modes identified through this analysis exhibit a direct correspondence with the field content of Soft-Collinear Effective Theory (SCET)~\cite{BurStw13lectures, BchBrgFrl15book, BurFlmLk00, BurPjlSwt02-1, BurPjlSwt02-2}. In particular, the on-shell-expansion region theorem guarantees that, under the wide-angle kinematic conditions of (\ref{eq:wideangle_onshell_kinematics}), only collinear and soft (termed ``ultrasoft'' in some literature) modes appear in the effective Lagrangian; equivalently, SCET$_{\rm I}$ and direct QCD yield identical predictions. Extending region analysis to the spacelike-collinear regime, where Glauber-mode momenta may contribute, should clarify the formulation of Glauber SCET~\cite{RstStw16} and deepen our understanding of the Regge limit, which has seen renewed interest recently~\cite{Falcioni:2021dgr,Caola:2021izf,Fadin:2021csi,Fadin:2023aen,Moult:2022lfy,Milloy:2022fns,Gao:2024fyz,Abreu:2024xoh}.

The methodology presented in this article applies specifically to amplitude calculations. Developing analogous region analysis techniques for phase-space integrals would have more direct relevance for LHC phenomenology. Initial progress in this direction has been made in Ref.~\citen{Smirnov:2024pbj}, where the small-mass asymptotic behavior of angular integrals is investigated from the geometric approach.

In recent years, there has been growing interests in the local subtraction of infrared singularities based on factorization, where subtraction terms are intimately related to the expansions in the MoR; see Refs.~\citen{AntsHndStmYangZeng20,AntsStm23,AntsKrlStmVkt24,KmsVcn24,Hnd25} for example. In these works, the explicit form of a local subtraction term is usually set up on a case-by-case basis, and knowledge of the MoR could pave the way toward a general, automated framework for local subtractions for general processes.

\section*{Acknowledgments}
This paper is based on several talks I gave during 2024–2025. I wish to express my gratitude to all participants for their valuable insights and stimulating discussions, including Charalampos Anastasiou, Thomas Becher, Martin Beneke, Claude Duhr, Einan Gardi, Xiao Liu, Alexander Penin, Ding Yu Shao, Chiara Signorile-Signorile, George Sterman, Iain Stewart, Robert Szafron, Gherardo Vita, Jian Wang, and many others. This work is supported by the Swiss National Science Foundation under the project funding scheme, grant number 10001706.

\section*{ORCID}
\noindent Yao Ma - \url{https://orcid.org/0000-0001-5067-2508}


\bibliographystyle{ws-ijmpa}
\bibliography{sample}







    
\end{document}